\documentclass[preprintnumbers]{revtex4}
\usepackage{amsfonts}
\usepackage{amssymb}
\usepackage{amsmath}
\usepackage{epsfig}
\usepackage{tabularx}
\usepackage{array}
\usepackage{cancel}
\allowdisplaybreaks
\begin{document}

\title{Gravitational decoupling of anisotropic stars in the Brans-Dicke theory}

\author{Kazuharu Bamba}
\email{bamba@sss.fukushima-u.ac.jp} \affiliation{Faculty of Symbiotic Systems Science, Fukushima University, Fukushima 960-1296, Japan}

\author{M. Z. Bhatti}
\email{mzaeem.math@pu.edu.pk} \affiliation{Department of
Mathematics, University of the Punjab, Quaid-i-Azam Campus,
Lahore-54590, Pakistan}

\author{Z. Yousaf}
\email{zeeshan.math@pu.edu.pk} \affiliation{Department of
Mathematics, University of the Punjab, Quaid-i-Azam Campus,
Lahore-54590, Pakistan}

\author{Z. Shoukat}
\email{zemishoukat@gmail.com} \affiliation{Department of
Mathematics, University of the Punjab, Quaid-i-Azam Campus,
Lahore-54590, Pakistan}

%%%%%
%%%%%
%\keywords{Relativistic stars; Alternative theories of gravity;
%Gravitation; Gravitational decoupling.}
%\pacs{04.40.Dg; 04.50.+h; 03.50.De; 04.40.-b.}
%%%%%
%%%%%

\begin{abstract}
Anisotropic spherically symmetric solutions within the framework of the Brans-Dicke theory are uncovered through a unique gravitational decoupling approach involving a minimal geometric transformation. This transformation effectively divides the Einstein field equations into two separate systems, resulting in the alteration of the radial metric component. The first system encompasses the influence of the seed source, derived from the metric functions of the isotropic Tolman IV solution. Meanwhile, the anisotropic source is subjected to two specific constraints in order to address the second system. By employing matching conditions to determine the unknown constants at the boundary of the stellar object, a comprehensive examination of the internal structure of stellar systems ensues.
This investigation delves into the impact of the decoupling parameter, the Brans-Dicke parameters, and a scalar field on the structural characteristics of anisotropic spherically symmetric spacetimes, all while considering the strong energy conditions.
\end{abstract}

\maketitle

\section{Introduction}

The theory of general relativity (GR) is one of the best theories of gravitational interaction. Even though its predictions were examined with extremely high precision, there remain unanswered issues due to which GR gets modified. Despite the great effort, it is difficult to introduce a single theory that encompasses all fundamental interactions, such as big-bang nucleosynthesis (the formation of primordial light elements) and cosmic microwave background radiation (CMBR). Schwarzschild \cite{schwarzschild1916sitzungsber} established the exact solution of Einstein field equations for the first time through a model with constant density. Since cosmic objects rarely have the same density across their interior regions, researchers perform different experiments with new approaches to find the better exact solution of Einstein field equations. Tolman \cite{tolman1939static} studied smooth matching conditions of interior space-time with the exterior one and computed a number of solutions in the presence of the cosmological constant $(\Lambda)$. Many astronomical investigations and present cosmological findings \cite{perlmutter1997measurements,perlmutter1998discovery,filippenko1998results,
caldwell2003phantom,tegmark2004cosmological,land2005examination,sherwin2011evidence} point out the expansion of our universe. In order to determine the composition of a stellar object in the context of any gravity theory, one may need to use an equation of state (EoS). The polytropic EoS serves as a base for the idea of polytropes. Polytropes in space-time with non-zero $\Lambda$ were thoroughly discussed in literature \cite{stuchlik2016general}. Some fascinating properties of polytropes for the system with spherical symmetric were studied in \cite{stuchlik2016general,novotny2017polytropic,stuchlik2017gravitational,hod2018lower,hod2018analytic}.

Anisotropy is incorporated into the structure of celestial objects by researchers by considering the radial and tangential ingredients of pressure which are unlikely to coincide. Lema\^{\i}tre \cite{lemaitre1933univers} considered a spherically symmetric metric, allowing the radial and tangential pressures to be different. He discussed that anisotropy occurs in low and high-density profiles owing to turning motion, phase-shift, and the existence of a magnetic field. The possible reasons for anisotropy in surface pressures may result from the existence of mixtures of various fluids, various types of phase transitions \cite{sokolov1980phase}, and other factors discussed by Lema\^{\i}tre. Ruderman \cite{ruderman1972pulsars} proposed that anisotropy can be generated by nuclear interactions in very dense systems, i.e., $\rho>10^{17}$kg m$^{-3}$. Numerous physically viable solutions were assessed in relation to relativistic theories of gravity in order to investigate the salient features of anisotropic matter distribution in \cite{bowers1974anisotropic,maurya2017anisotropic,
matondo2018relativistic}.
It is observed that local anisotropies are crucial to understand how compact structures behave. However, determining the optimal approach to incorporate anisotropy into the stellar distribution remains an important question. Einstein field equations are quite challenging to solve due to their complex mathematical structure, unless certain constraints are put in place. The problem of solving Einstein field equations is slightly simpler when the stellar distribution is isotropic rather than anisotropic. This is due to the fact that there are only four unknowns in isotropic configuration $(\rho, p, e^{\nu}, e^{\zeta})$, whereas there are five unknowns in the anisotropic case $(\rho, p_{r}, p_{t}, e^{\nu}, e^{\zeta})$. Here, $\rho$ is energy density, $p$ is isotropic pressure, while $p_{r}$ stands for radial pressure and $p_{t}$ denotes tangential pressure and $(e^{\nu}, e^{\zeta})$ indicate the metric functions. It is observed that in an anisotropic case, some additional components are added to the matter distribution due to which the number of unknowns grows which makes it more challenging to solve the Einstein field equations analytically. In this regard, such analytical solutions developed via gravitational decoupling (GD) with minimal geometric deformation (MGD) method in both cosmology and astrophysics \cite{ovalle2017decoupling,ovalle2018anisotropic}.

Multiple methods were studied to investigate important properties of self-gravitating objects including such as the phenomenon of stability and hydrodynamic equilibrium, the upper limit of the mass-to-radius ratio, the upper limit of superficial redshift, and dynamics of matter content under energy conditions \cite{buchdahl1959general}. One of these techniques is the MGD approach, which was initially intended as an optional means of deforming Schwarzschild space-time in the framework of the Randall-Sundrum braneworld \cite{randall1999large,randall1999alternative}. Recently, there has been a lot of interest in developing novel analytic and anisotropic solutions for Einstein field equations, which is a difficult task as Einstein field equations are non-linear and difficult to handle. In this way, the method of MGD to gain new models representing relativistic objects with well-determined characteristics has been proposed \cite{ovalle2008searching}. For a compact spherical distribution, the analytical solution of an anisotropic fluid as well as the braneworld model of Tolman IV solution have been found \cite{ovalle2013tolman}. Two essential components are required, one is the dimensionless coupling constant $``\alpha"$ to incorporate an extra source into the stress-energy tensor of the seed solution. The second one is the MGD method on the metric potentials (often on the radial component of the metric) in the context of the braneworld model. If the seed solution is assumed to be anisotropic, the inclusion of this additional component combined with a static and spherically symmetric system gives rise to complex simultaneous equations. The MGD technique separates Einstein field equations into two systems, namely the ``Einstein system" and ``quasi-Einstein system", which in comparison to the original system are easy to solve. At this point, a few observations are appropriate, firstly, the decoupled systems satisfy Bianchi Identities and secondly, the extra source may be a scalar, vector, or tensor field \cite{ovalle2013role,casadio2014black,casadio2015classical,casadio2015minimal,ovalle2018simple}. Moreover, a number of interesting results on the solutions of black holes with 2+1 and 3+1 decomposition were obtained in \cite{ovalle2018black,contreras2018minimal,contreras2019gravitational,rincon2020anisotropic}. Additionally, the solutions of new hairy black holes have just been explored \cite{ovalle2021hairy}, and a mechanism is created as well to turn any non-rotating black hole into a rotational one \cite{contreras2021gravitational,ovalle2021kerr}.

When weak gravitational forces are at work, the hypothesis in GR has effectively aligned with many tests carried out within the solar system, demonstrating its success in cosmology. To get more accurate and dependable results, this theory may need to be modified when dealing with high gravitational fields or while being observed on a big scale. These changes may be very important in explaining the phenomena of accelerated expansion. These modifications are termed modified theories (see, for instance \cite{odintsov2006introduction,nojiri2007introduction, padmanabhan2008dark,clifton2012modified,capozziello2011extended,nojiri2011unified,sotiriou2010f,
de2010f,joyce2015beyond,nojiri2017modified})

%The dynamical stability of compact systems with an anisotropic environment
%has been investigated in $f(R, T)$ gravity \cite{bhatti2022dynamical,bhatti2022dynamicala,bhatti2022study}. They used perturbation techniques to derive the collapse equation, from the conservation equations, field equations, modified gravity's extra curvature terms which appear due to specific $f(R, T)$ model. They also obtained certain meaningful constraints for the adiabatic index in both the Newtonian and Post Newtonian regimes to ensure the stability of the celestial self-gravitating configuration.
Many modified theories of gravity \cite{wagoner1970scalar,lovelock1972four,ford1989inflation,alcaraz2003limits,nojiri2006modified, arai2023cosmological} are taken into account by changing the Einstein-Hilbert action that is frequently used to study both the existence of dark energy and dark matter as well as the mystery of universe rapid expansion. Geometrical representation and scalar-tensor representation of $f(G,T)$ gravity have been presented to establish novel junction condition \cite{bhatti2023novel}.

%Sharif \emph{ et al.} \cite{sharif2020anisotropic,sharif2021compact} studied different solutions for self-gravitating systems in modified gravity theories by using gravitational decoupling approach to transform metric potentials and separate the field equations. They also incorporated Krori-Barua spacetime metrics to expand the isotropic solution. Two anisotropic solutions are found by imposing physical constraints. They analyzed the physical viability of the solutions, equilibrium, and stability of considered system. They observed certain extensions must meet for specific decoupling values.
%The compact objects alternative to black hole, i.e., the so-called gravastar, has been studied under the influence of modified gravity \cite{yousaf2020construction,yousaf2020gravastars}.

Several researchers are interested in exploring the gravitational collapse phenomena because it is a prominent case in a strong-field regime \cite{kwon1986stability,shibata1994scalar,harada1997scalar}.
%Dirac \cite{dirac1937cosmological,dirac1938nature} noticed that a connection between dimensionless combinations of $\Lambda$ and physical constants naturally develops, when one of the constants is allowed to change through cosmic time scales. Dirac's choice made the gravitational constant time dependent while leaving the other basic constants unaffected.
Jordan \cite{jordan1938empirischen} developed a full gravitational theory that gave the title of a gravitational scalar field to gravitational constant. Brans and Dicke \cite{brans1961mach} developed a scalar-tensor field theory named the Brans-Dicke (BD) theory obtained by substituting a time modifying constant $G(t)$ and with the help of a scalar field $(\Phi)$ having interaction along with the geometry. Additionally, the well-known scalar field coupling constant or parameter ($\omega_{BD}$) of the BD theory is a constant that can be adjusted to get the desired outcomes in the Jordan frame. It is assumed that the $\Phi$ is reciprocal of the dynamical gravitational constant, i.e., $G(t)=\frac{1}{\Phi(t)}$. The test particles travel along geodesics according to the BD theory, they consequently obey the weak equivalence principle, which states that the gravitational mass and inertial mass are equivalent. Mach principle, agreement with the weak equivalence principle, and Dirac's large number hypothesis are the main ingredients of the BD theory. This theory includes a metric tensor and a scalar field that describes gravity.

%Dehnen and Obreg\'{o}n \cite{dehnen1972exact} studied the exact solutions of BD cosmic equations, which have no correspondence in GR even for the bigger values of the parameter $\omega_{BD}$. The BD theory of gravitation demonstrated that the inflationary era holds for lower values of $\omega_{BD}$ \cite{weinberg1989some}.
%Rama \emph{et al.} \cite{rama1996short} studied through weak-field experiments that the general dimensionless coupling constant $\omega_{BD}$ should be large as $\omega_{BD}\geq 4\times10^{4}$.
%Romero and  Barros \cite{romero1993brans} studied the BD exterior solutions with the existence of $\Lambda$ from the perspective of dynamical system.
The large value of $\Phi$ describes the fast expansion of the universe and is found by a recent study in cosmology, including the redshift and distance-luminosity connection of type Ia Supernovae \cite{riess1998observational}. The evidence for various cosmic concerns, including the late behavior of the universe, cosmic acceleration, and the inflation issue, etc. are also supported by the BD theory \cite{banerjee2001cosmic}. %The scalar-tensor field theory is the most well-known and straightforward modified theory of gravity \cite{fujii2003scalar}.
This theory has drawn interest in recent years due to its precision in describing the early inflationary era and late-time expansion of the cosmos. Several authors studied the Friedmann-Lema\^{\i}tre-Robertson-Walker model in the context of the BD theory \cite{sen2001dissipative,karimkhani2019hubble,singh2021friedmann}.
%Yousaf \emph{et al.} \cite{yousaf2023dynamics} studied the effects of magnetic fields of non-linear electrodynamics in chameleonic BD theory under the existence of anisotropic spherical fluid.

The main purpose of this work is to extend \cite{ovalle2022energy} in the BD theory. The paper is organized as follows. In Sec. II, the appropriate BD theory for the GD formalism is discussed. In Sec. III, the MGD technique to a spherically symmetric geometry filled with two sources in the BD theory is introduced. In Sec. IV, junction conditions are established to match an outside Schwarzschild line element with the inside solution. In Sec. V, we studied mathematical and physical solutions to the modified field equations using the MGD method with constraints applied to matter density and radial pressure for anisotropy in the setting of the BD theory. In Sec. VI, the physical characteristics of an anisotropic stellar structure with the help of a polytropic equation of state are provided. Finally, in Sec. VII, the important results including discussions are summarized and further insights and outlooks are also described.

\section{Brans-Dicke theory and modified field equations}

The four-dimensional modified gravitational action of the BD theory is given as \cite{nordtvedt1970post,wagoner1970scalar}
\begin{align}\label{1}
S^{(BD)}=\frac{1}{2}\int(\Phi R -\frac{\omega_{BD}}{\Phi}g^{\sigma\nu}\nabla_{\sigma}\Phi\nabla_{\nu}\Phi)\sqrt{-g} d^4x+\int \mathcal{L}_{M}\sqrt{-g} d^4x+\int \mathcal{L}_{X}\sqrt{-g} d^4x,
\end{align}
where $R,~g$, $\omega_{BD}$, $\mathcal{L}_{M}$ and $\mathcal{L}_{X}$ denote Ricci scalar, determinant of metric tensor $g_{\sigma\nu}$, generic dimensionless parameter of the BD theory, Lagrangian density for the matter and new gravitational source not defined in GR, respectively. The stress-energy tensor, connected with two Lagrangian density functions $\mathcal{L}_{M}$ and $\mathcal{L}_{X}$, is given as
\begin{align*}
T_{\sigma\nu}=-\frac{2}{\sqrt{-g}}\frac{\delta(\sqrt{-g}\mathcal{L}_{M})}{\delta g_{\sigma\nu}}=g_{\sigma\nu}\mathcal{L}_{M}-2\frac{\partial \mathcal{L}_{M}}{\partial g^{\sigma\nu}},\\\nonumber
\vartheta_{\sigma\nu}=-\frac{2}{\sqrt{-g}}\frac{\delta(\sqrt{-g}\mathcal{L}_{X})}{\delta g_{\sigma\nu}}=g_{\sigma\nu}\mathcal{L}_{X}-2\frac{\partial \mathcal{L}_{X}}{\partial g^{\sigma\nu}},
\end{align*}
which shows Lagrangian density is simply dependent on the metric tensor $g_{\sigma\nu}$. We consider the modified field equations in framework of the BD theory and GD method which takes the following form
\begin{align}\label{2}
G_{\sigma\nu}\equiv R_{\sigma\nu}-\frac{1}{2}R~g_{\sigma\nu}=\frac{1}{\Phi} \tilde{T}_{\sigma\nu},
\end{align}
along with total stress-energy tensor as
\begin{align*}
\tilde{T}_{\sigma\nu}={T_{\sigma\nu}}^{(m)}+{T_{\sigma\nu}}^{(\Phi)}+\vartheta_{\sigma\nu},
\end{align*}
here, ${T_{\sigma\nu}}^{(m)}$ is usual matter associated to known solution, ${T_{\sigma\nu}}^{(\Phi)}$ is stress-energy tensor of the BD theory, and $\vartheta_{\sigma\nu}$ is new gravitational source not supported by GR.
\begin{align}\label{3}
\Box\Phi=\frac{T^{(m)}+\vartheta}{3+2\omega_{BD}},
\end{align}
where, $\Box\equiv\nabla^{\eta}\nabla_{\eta}$ is d'Alembertian operator. The above equation explains evolution of scalar field, where $T^{(m)}$ and $\vartheta$ represent the trace of $T_{\sigma\nu}$ and $\vartheta_{\sigma\nu}$, respectively. They are defined as
\begin{align*}
T^{(m)}= g^{\sigma\nu} T_{\sigma\nu},\quad \vartheta= g^{\sigma\nu} \vartheta_{\sigma\nu}.
\end{align*}
Since, the total source must be covariantly conserved of the system, which can be obtained by using Bianchi Identity as
\begin{align}\label{4}
\nabla_{\sigma} \tilde{T}^{\sigma\nu}=0.
\end{align}
The interior line element of spherically symmetric compact object is symbolized in Schwarzschild coordinates ($t,~r,$ $\theta$, $\phi$) which can be written in the form
\begin{align}\label{5}
ds^2=e^{\nu(r)}dt^2-e^{\zeta(r)}dr^2-r^2 d\theta^2-r^2sin^2\theta d\phi^2,
\end{align}
where $e^{\nu(r)}$ and $e^{\zeta(r)}$ denote the metric functions. Furthermore, we employ the signature $(+,-,-,-)$ and relativistic units, i.e, $c=1, 8\pi=1$. In this regard, we take an anisotropic distribution of fluid as
\begin{align}\label{6}
{T_{\sigma\nu}}^{(m)}=(\rho+p_{t})\mathcal{U}_{\sigma}\mathcal{U}_{\nu}-p_{t}g_{\sigma\nu}+(p_{r}-p_{t})\mathcal{X}_{\sigma}\mathcal{X}_{\nu},
\end{align}
where, $\mathcal{U}_{\sigma}$ and $\mathcal{X}_{\sigma}$ denote the four-velocity and unit four-vector along the radial direction, respectively. However, $\rho$, $p_{r}$, and $p_{t}$ are the energy density, radial pressure, and tangential pressure, respectively. The four-velocity and unit four-vector are calculated by using the expression $\mathcal{U}^{\sigma}=e^{\frac{-\nu}{2}}\delta_{0}^{\sigma}$ and $\mathcal{X}^{\sigma}=e^{\frac{-\zeta}{2}}\delta_{1}^{\sigma}$, respectively, which satisfy the following conditions under comoving frame
\begin{align*}
\mathcal{U}_{\sigma}\mathcal{U}^{\sigma}=1,\quad\mathcal{X}^{\sigma}\mathcal{U}_{\sigma}=0,\quad\mathcal{X}^{\sigma}\mathcal{X}_{\sigma}=-1.
\end{align*}
The canonical expression of Eq. \eqref{3} can be written in the following form
\begin{align*}
{T_{\sigma\nu}}^{(m)}=\rho~\mathcal{U}_{\sigma}\mathcal{U}{\nu}+\Pi_{\sigma\nu}+p h_{\sigma\nu},
\end{align*}
with
\begin{align*}\nonumber
&p=\frac{2p_{t}+p_{r}}{3},\quad h_{\sigma\nu}=g_{\sigma\nu}+\mathcal{U}_{\sigma}\mathcal{U}_{\nu},\quad
\Pi_{\sigma\nu}=\Pi\bigg(\mathcal{X}_{\sigma}\mathcal{X}_{\nu}-\frac{1}{3}h_{\sigma\nu}\bigg),\quad\Pi=p_{r}-p_{t},
\end{align*}
where $h_{\sigma\nu}$ is projection tensor, $\Pi_{\sigma\nu}$ anisotropic tensor, and $\Pi$ anisotropic factor, respectively. The stress-energy tensor of the scalar field is expressed as
\begin{align}\label{7}
T^{(\Phi)}_{\sigma\nu}=\Phi_{,\sigma;\nu}-\Box\Phi g_{\sigma\nu}+\frac{\omega_{BD}}{\Phi}\bigg(\Phi_{,\sigma}\Phi_{,\nu}
-\frac{1}{2}g_{\sigma\nu}\Phi^{,\eta}\Phi_{,\eta}\bigg),
\end{align}
which describes energy related to the scalar field. To make ``$\omega_{BD}$" dimensionless, the factor $``\Phi"$ is added to the denominator of the last term in the above equation. The matter variables and the metric tensor $g_{\sigma\nu}$ along with the scalar field $\Phi$ characterize the dynamics of the gravitational field. The modified field equations are evaluated by using Eqs. \eqref{2} and \eqref{5} which provide the following expressions
\begin{align}\label{8}
\frac{1}{\Phi}(\rho+\rho^{(\Phi)}+\mathcal{\varepsilon})&=\frac{1}{r^{2}}-e^{-\zeta}\bigg(\frac{1}{r^{2}}
-\frac{\zeta'}{r}\bigg),\\\label{9}
\frac{1}{\Phi}(p_{r}+p_{r}^{(\Phi)}+\mathcal{P}_r)&=-\frac{1}{r^{2}}
+e^{-\zeta}\bigg(\frac{1}{r^{2}}+\frac{\nu'}{r}\bigg),\\\label{10}
\frac{1}{\Phi}(p_{t}+p_{t}^{(\Phi)}+\mathcal{P}_t)&=
\frac{e^{-\zeta}}{4}\bigg(2\nu''+\nu'^2-\zeta'\nu'+2\frac{\nu'-\zeta'}{r}\bigg),
\end{align}
where prime notation indicates the derivative with respect to $``r"$. The above set of Eqs. \eqref{8}-\eqref{10} describe the relation between a gravitational field and an imperfect fluid. Moreover, the constituents of the BD theory $\rho^{(\Phi)}, p_{r}^{(\Phi)}$, and $p_{t}^{(\Phi)}$ can be given in terms of $\nu$ and $\zeta$ such that
\begin{align}\label{11}
\rho^{(\Phi)}&=-e^{-\zeta}\bigg[\Phi''+\Phi'\bigg(\frac{2}{r}-\frac{\zeta'}{2}\bigg)
+\frac{\omega_{BD}\Phi'^{2}}{2\Phi}\bigg],\\\label{12}
p_{r}^{(\Phi)}&=-e^{-\zeta}\bigg[\Phi'\bigg(\frac{2}{r}+\frac{\nu'}{2}\bigg)
-\frac{\omega_{BD}\Phi'^{2}}{2\Phi}\bigg],\\\label{13}
p_{t}^{(\Phi)}&=-e^{-\zeta}\bigg[\Phi''+\Phi'\bigg(\frac{1}{r}+\frac{\nu'}{2}
-\frac{\zeta'}{2}\bigg)+\frac{\omega_{BD}\Phi'^{2}}{2\Phi}\bigg].
\end{align}
%The generally covariant d'Alembert is defined to be the covariant divergence of $\Phi^{,\alpha}$
The equation for evolution of $\Phi$ can be obtained by using Eq. \eqref{3} with metric \eqref{5} which takes the following form
\begin{align}\label{14}
\Box\Phi=-e^{-\zeta}\bigg[\Phi''+\bigg(\frac{2}{r}+\frac{\nu'}{2}-
\frac{\zeta'}{2}\bigg)\Phi'\bigg].
\end{align}
We consider a spherically symmetric system which implies that $G_{2}^{2}=G_{3}^{3}$, therefore, we take $\tilde{T}_{2}^{2}=\tilde{T}_{3}^{3}$. Now, Eqs. \eqref{8}-\eqref{10} can be used to find the expression for $\tilde{\rho}$, $\tilde{p_r}$, and $\tilde{p_t}$ such that
\begin{align}\label{15}
\tilde{\rho}=\rho+\rho^{(\Phi)}+\mathcal{\varepsilon},\\\label{16}
\tilde{p_r}=p_{r}+p_{r}^{(\Phi)}+\mathcal{P}_r,\\\label{17}
\tilde{p_t}=p_{t}+p_{t}^{(\Phi)}+\mathcal{P}_t,
\end{align}
where $\tilde{\rho}$, $\tilde{p_r}$, and $\tilde{p_t}$ represent the effective density, effective radial pressure, and effective tangential pressure, respectively. From Eqs. \eqref{15}-\eqref{17}, we can observe that metric \eqref{5} is filled by
\begin{align*}
{T_{\sigma}^{\nu}}^{(m)}&= diag[\rho,-p_{r},-p_{t},-p_{t}],\\
{T_{\sigma}^{\nu}}^{(\Phi)}&= diag[\rho^{(\Phi)},-{p_{r}}^{(\Phi)},-{p_{t}}^{(\Phi)},-{p_{t}}^{(\Phi)}],\\
\vartheta_{\sigma}^{\nu}&= diag[\varepsilon,-{\mathcal{P}}_{r},-{\mathcal{P}}_{t},-{\mathcal{P}}_{t}].
\end{align*}
The anisotropy develops inside the stellar distribution due to the presence of effective radial and effective tangential pressure components, which turn out
\begin{align}\label{18}
\tilde{\Pi}=\tilde{p}_{t}-\tilde{p}_{r}=({p}_{t}-{p}_{r})+({{p}_{t}}^{(\Phi)}-{{p}_{r}}^{(\Phi)})+(\mathcal{{P}}_{t}-\mathcal{{P}}_{r}),
\end{align}
here
\begin{align*}
p_{t}^{(\Phi)}-p_{r}^{(\Phi)}=-e^{-\zeta}\bigg[\Phi''-\bigg(\frac{1}{r}+\frac{\zeta'}{2}\bigg)\Phi'+\frac{\omega_{BD}\Phi'^{2}}{\Phi}\bigg],
\end{align*}
where the degree of anisotropy is determined by different values of $\omega_{BD}$ parameter. Anisotropy vanishes if we have a small value of $\omega_{BD}$, similarly, it becomes more prominent if we have larger values of $\omega_{BD}$. From Eq. \eqref{18}, we can see that if the scalar field $\Phi$ is constant with respect to the radial direction, and no anisotropy due to the scalar field. It is clear that Eqs. \eqref{8}-\eqref{13} treated as an anisotropy of the fluid. Also with a new gravitational source, we can observe more effects of anisotropy. In the next section, we apply the GD approach to the system of modified field equations \eqref{8}--\eqref{10} to solve them.

\section{Gravitational Decoupling}

In this section, we precisely discuss the GD for spherically symmetric metric (for details see \cite{ovalle2019decoupling}). The GD approach through MGD method \cite{ovalle2020beyond,arias2020extra,tello2020class,maurya2021mgd,maurya2022gravitationally} is attractive for many reasons. Some significant applications include, firstly, connecting gravitational sources, enabling the extension of known solutions in the Einstein field equations into more intricate domains. Secondly, the separation of gravitational sources systematically reduces (decouples) a complex stress-energy tensor into simpler components. Lastly, it explores solutions within gravitational theories that go beyond Einstein's framework, among numerous other applications. We consider the solution of the Eq. \eqref{2} without a new gravitational source which implies
\begin{align}\label{19}
\tilde{~T_{\sigma}^{\nu}}= {T_{\sigma}^{\nu}}^{(m)}+{T_{\sigma}^{\nu}}^{(\Phi)}+\cancelto {0}{\vartheta_{\sigma}^{\nu}}.
\end{align}
We can also define
\begin{align*}
{T_{\sigma}^{\nu}}^{(eff)}={T_{\sigma}^{\nu}}^{(m)}+{T_{\sigma}^{\nu}}^{(\Phi)m},\quad {\vartheta_{\sigma}^{\nu}}^{(eff)}=\vartheta_{\sigma}^{\nu}+{T_{\sigma}^{\nu}}^{(\Phi)\vartheta}.
\end{align*}
and one can write for space-time \eqref{5} as
\begin{align}\label{20}
ds^2=e^{\chi(r)}dt^2-e^{\sigma(r)}dr^2-r^2 d\theta^2-r^2sin^2\theta d\phi^2.
\end{align}
The Misner-Sharp mass function for the un-deformed metric \eqref{20} computed as
\begin{align}\label{21}
e^{-\sigma(r)}\equiv 1-\frac{1}{r}\int_{0}^{r}x^{2}\bigg({\rho}^{(\Phi)}+\rho\bigg)dx=1-\frac{2 m(r)}{r}.
\end{align}
The above expression shows the role of scalar field and energy density within the surface of radius $r$. In this sense, results including the source $\vartheta_{\sigma}^{\nu}$ may be observed as a geometric deformation of the Eq. \eqref{20}, namely
\begin{align}\label{22}
\chi\rightarrow \nu&=\chi+g^{*},\\\label{23}
e^{-\sigma}\rightarrow e^{-\zeta}&=e^{-\sigma}+f^{*},
\end{align}
where $f^{*}, g^{*}$ are the decoupling functions for radial and temporal metric components, respectively. Note that transformations in the Eqs. \eqref{22} and \eqref{23} represent change in the space-time geometry \eqref{20}. With the help of these transformations, Eqs. \eqref{8}--\eqref{10} are separated into two subsets, one set corresponding to the original stress-energy tensor whose metric is given by Eq. \eqref{20}. Consequently, we compute
\begin{align}\label{24}
\rho&=\frac{1}{r^{2}}-e^{-\sigma}\bigg(\frac{1}{r^{2}}-\frac{\sigma'}{r}\bigg)
-e^{-\sigma}\bigg(\Phi''+\frac{2\Phi'}{r}-\frac{\Phi'\sigma'}{2}+\frac{\omega_{BD}\Phi'^{2}}{2\Phi}\bigg),\\\label{25}
p_{r}&=-\frac{1}{r^{2}}+e^{-\sigma}\bigg(\frac{1}{r^{2}}+\frac{\chi'}{r}\bigg)
+e^{-\sigma}\bigg(\frac{2\Phi'}{r}+\frac{\chi'\Phi'}{2}-\frac{\omega_{BD}\Phi'^{2}}{2\Phi}\bigg),\\\label{26}
p_{t}&=\frac{e^{-\sigma}}{4}\bigg(2\chi''+\chi'^2-\sigma'\chi'+2\frac{\chi'-\sigma'}{r}\bigg)+e^{-\sigma}(\Phi''+\frac{\Phi'}{r}+\frac{\chi'\Phi'}{2}+\frac{\omega_{BD}\Phi'^{2}}{2\Phi}-\frac{\Phi'\sigma'}{2}).
\end{align}
Another set corresponds to a new gravitational source $\vartheta_{\sigma}^{\nu}$, reads as
\begin{align}\label{27}
\varepsilon&=-\frac{f^{*}}{r^{2}}-\frac{{f}'^{*}}{r}-f^{*}\Phi'' -\frac{2\Phi'f^{*}}{r}-\frac{{f}'^{*}\Phi'}{2}-\frac{\omega_{BD}\Phi'^{2}f^{*}}{2\Phi},\\\label{28}
\mathcal{P}_{r}-Z_{1}&=f^{*}\bigg(\frac{1}{r^{2}}+\frac{\nu'}{r}\bigg)+f^{*}\bigg(\frac{2\Phi'}{r}
+\frac{\nu'\Phi'}{2}-\frac{\omega_{BD}\Phi'^{2}}{2\Phi}\bigg),\\\label{29}
\mathcal{P}_{t}-Z_{2}&=\frac{f^{*}}{4}\bigg[2\nu''+\nu'^{2}+\frac{2\nu'}{r}
+4\bigg(\Phi''+\frac{\Phi'}{r}+\frac{\nu'\Phi'}{2}+
\frac{\omega_{BD}\Phi'^{2}}{2\Phi}\bigg)\bigg]+
\frac{{f}'^{*}}{4}\bigg(\nu'+\frac{2}{r}+2\Phi'\bigg),
\end{align}
where
\begin{align*}
Z_{1}&=\frac{e^{-\sigma}{{g}^{*}}'}{r}+\frac{e^{-\sigma}{g}'^{*}\Phi'}{2},\quad
Z_{2}=\frac{e^{-\sigma}}{4}\bigg(2{g}''^{*}+{g}'^{*2}+\frac{2{g}'^{*}}{r}+2\chi'{g}'^{*}-\sigma'{g}'^{*}
+2{g}'^{*}\Phi'\bigg).
\end{align*}
We observe that the tensor ``$\vartheta_{\sigma}^{\nu}$" disappears when the deformations disappear, i.e., $f^{*} = g^{*} = 0$. Equations \eqref{27}-\eqref{29} are termed as the quasi-Einstein system that consists of two deformation functions $f^{*}$ and $g^{*}$ for radial and temporal metric components, respectively. If we want to take only the radial metric component, then we have to take $g^{*}$=0. This special case where we take $g^{*}$=0 comes under gravitational decoupling through the minimal geometric deformation (MGD) technique, in which $\vartheta_{\sigma}^{\nu}$ is only determined by $f^{*}$.
In other words, the new gravitational source leads to a modification of the radial component of the metric according to Eq. \eqref{23}, but the temporal component of the metric stays unaltered \cite{ovalle2017decoupling}. In this situation, the dynamical equation may be computed by using Eq. \eqref{4} as
\begin{align}\nonumber
&\bigg[{T_{1}^{1}}'^{(eff)}-\frac{\chi'}{2}\bigg({T_{0}^{0}}^{(eff)}-{T_{1}^{1}}^{(eff)}\bigg)
-\frac{2}{r}\bigg({T_{2}^{2}}^{(eff)}-({T_{1}^{1}}^{(eff)}\bigg)\bigg]
-\frac{{g}'^{*}}{2}\bigg({T_{0}^{0}}^{(eff)}-{T_{1}^{1}}^{(eff)}\bigg)
+\bigg[{\vartheta_{1}^{1}}'^{(eff)}-\\\label{30}
&\frac{\nu'}{2}\bigg({\vartheta_{0}^{0}}^{(eff)}-{\vartheta_{1}^{1}}^{(eff)}\bigg)
-\frac{2}{r}\bigg({\vartheta_{2}^{2}}^{(eff)}
-{\vartheta_{1}^{1}}^{(eff)}\bigg)\bigg]=0,
\end{align}
in which the term included in square brackets is the divergence of the usual stress-energy tensor and corresponds to a known seed source calculated with $\nabla^{(\chi, \sigma)}$ as a covariant derivative for the line element \eqref{20} and is a combination of the modified field equations \eqref{24}--\eqref{26}. The last term in square brackets corresponds to a new gravitational source in which the line element \eqref{5} is used to calculate the divergence, as
\begin{align}\label{31}
\nabla_{\mu}^{(\chi,\sigma)}{(T_{\nu}^{\mu}}^{(eff)})-\frac{{g}'{*}}{2}\bigg[{T_{0}^{0}}^{(eff)}-{T_{1}^{1}}^{(eff)}\bigg]
+\nabla_{\mu}{\vartheta_{\nu}^{\mu}}^{(eff)}=0.
\end{align}
The above equation follows that the stress-energy tensor of matter must have a vanishing covariant divergence, the source ${T_{\sigma}^{\nu}}^{(eff)}$ is conserved with respect to metric \eqref{20}. Thus, by applying Bianchi identities, we obtain the form
\begin{align}\label{32}
\nabla_{\nu}^{(\chi,\sigma)}{T_{\sigma}^{\nu}}^{(eff)}=0.
\end{align}
It is noted that we have
\begin{align}\label{33}
\nabla_{\nu}{T_{\sigma}^{\nu}}^{(eff)}=\nabla_{\nu}^{(\chi,\sigma)}{T_{\sigma}^{\nu}}^{(eff)}
-\frac{{g}'^{*}}{2}\bigg[{T_{0}^{0}}^{(eff)}-{T_{1}^{1}}^{(eff)}\bigg]\delta_{\sigma}^{1}.
\end{align}
By substituting Eq. \eqref{32} into Eq. \eqref{33}, we find
\begin{align}\label{34}
\nabla_{\nu}{T_{\sigma}^{\nu}}^{(eff)}=-\frac{{g}'^{*}}{2}\bigg[{T_{0}^{0}}^{(eff)}-{T_{1}^{1}}^{(eff)}
\bigg]\delta_{\sigma}^{1},
\end{align}
where, the values of ${T_{0}^{0}}^{(eff)}$, ${T_{1}^{1}}^{(eff)}$, and ${T_{2}^{2}}^{(eff)}$ are given in Appendix A (due to the long expressions). The divergence of the gravitational source ${\vartheta_{\sigma}^{\nu}}^{(eff)}$ turn out
\begin{align}\label{35}
\nabla_{\nu}{\vartheta_{\sigma}^{\nu}}^{(eff)}=\frac{{g}'^{*}}{2}\bigg[{T_{0}^{0}}^{(eff)}-{T_{1}^{1}}^
{(eff)}\bigg]\delta_{\sigma}^{1},
\end{align}
along with
\begin{align*}
&{\vartheta_{0}^{0}}^{(eff)}=\vartheta_{0}^{0}+f^{*}\bigg(\Phi''+\frac{2\Phi'}{r}
+\frac{\omega_{BD}\Phi'^{2}}{2\Phi}\bigg)+\frac{{f}'^{*}\Phi'}{2},\\
&{\vartheta_{1}^{1}}^{(eff)}=\vartheta_{1}^{1}+f^{*}\bigg(\frac{2\Phi'}{r}
+\frac{\nu'\Phi'}{2}-\frac{\omega_{BD}\Phi'^{2}}{2\Phi}\bigg)+\frac{{g}'^{*}\Phi'e^{-\sigma}}{2},\\
&{\vartheta_{2}^{2}}^{(eff)}=\vartheta_{2}^{2}+f^{*}\bigg(\Phi''+\frac{\Phi'}{r}
+\frac{\Phi'\nu'}{2}+\frac{\omega_{BD}\Phi'^{2}}{2\Phi}\bigg)+\frac{{f}'^{*}\Phi'}{2}
+\frac{{g}'^{*}\Phi'e^{-\sigma}}{2}.
\end{align*}
The expressions in Eqs. \eqref{34} and \eqref{35} describe both sources ${T_{\sigma}^{\nu}}^{(m)}$ and $\vartheta_{\sigma}^{\nu}$ which may be successfully decoupled provided that there is an energy transfer between them. We observe that the GD approach can effectively decouple both sources ${T_{\sigma}^{\nu}}^{(m)}$ and $\vartheta_{\sigma}^{\nu}$ which is especially impressive since there is no need for any perturbative expansion in $f^{*}$ or $g^{*}$ \cite{ovalle2020beyond}. Thus, the information about the stress-energy exchange system could be obtained by using Eqs. \eqref{34} and \eqref{35}. Therefore,
\begin{align}\label{36}
\Delta E=\frac{{g}'^{*}}{2}\bigg\{(\rho+\rho^{(\Phi)})+(p_{r}+p_{r}^{(\Phi)})\bigg\}.
\end{align}
This shows that energy exchange depends upon the linear combination of matter variables and deformation factor $g'$ within the context of the BD theory. We can write Eq. \eqref{36} with the help of Eqs. \eqref{24}--\eqref{26} in terms of pure geometric functions as
\begin{align}\label{37}
\Delta E=\frac{{g}'^{*}e^{-\sigma}}{2r}(\sigma'+\chi').
\end{align}
The expression \eqref{36} shows that ${g}'^{*}>0$ which results in $\Delta E > 0$. Therefore, according to Eqs. \eqref{34} and \eqref{35}, we have $\nabla_{\nu}{\vartheta_{\sigma}^{\nu}}^{(eff)}>0$, which indicates the source $\vartheta_{\sigma}^{\nu}$ with dark source terms of the BD theory is giving energy to the environment. Similarly, $\vartheta_{\sigma}^{\nu}$ gains energy from the environment provided ${g}'^{*}<0$.

\section{Junction Conditions}

The matching of the internal and external geometries at the star surface is a significant aspect in the study of stellar distributions \cite{israel1967nuovo}. We follow the approach given by Israel-Darmois \cite{darmois1927memorial,israel1966singular} for the junction conditions. The interior metric is described by Eq. \eqref{5} while
% given as
%\begin{align}\label{38}
%ds^{2}=e^{\varpi(r)}dt^2-e^{-\varrho(r)}dr^2-r^2\sin^2\theta d\phi^2,
%\end{align}
%where $e^{-\varrho(r)}=1-\frac{2\tilde{m}(r)}{r}$ and
interior mass function $\tilde{m}(r)$ is given by
\begin{align}\label{39}
\tilde{m}(r)=m(r)-\frac{r}{2}f^{*}(r),
\end{align}
where  $m(r)$ is given in Eq. \eqref{21} and $f^{*}(r)$ is the deformation factor as given in transformation Eq. \eqref{23}. For the exterior geometry, we select Schwarzschild metric having expression in the following form
\begin{align}\label{40}
ds^{2}=\bigg(1-\frac{2\mathcal{M}}{r}\bigg)dt^2-\bigg(1-\frac{2\mathcal{M}}{r}\bigg)^{-1}dr^2-r^2\sin^2\theta d\phi^2,
\end{align}
where $\mathcal{M}$ represents total mass. The following conditions should be fulfilled to have continuity and smoothness of geometry at the star surface. Therefore,
\begin{align*}
(ds^{2}_{-})_\Sigma=(ds^{2}_{+})_\Sigma;~~~({K_{\sigma\nu}}_{-})_\Sigma=({K_{\sigma\nu}}_{+})_\Sigma;~~~
(\Phi(r)_{-})_\Sigma=(\Phi(r)_{+})_\Sigma;~~~(\Phi'(r)_{-})_\Sigma=(\Phi'(r)_{+})_\Sigma.
\end{align*}
Here, $K_{\sigma\nu}$ indicates curvature tensor while subscripts ``+" and ``-" denote the exterior and interior space-time, respectively. The continuity of first fundamental form can be given as
\begin{align}\label{41}
e^{\varpi(R)}=1-\frac{2\mathcal{M}}{r},\quad\quad&\quad\quad\quad\quad\quad\quad\quad e^{-\varrho(R)}=1-\frac{2\mathcal{M}}{r}.
\end{align}
Similarly, we obtain the second fundamental form by taking the Israel-Darmois junction condition at the surface such that
\begin{align}\label{42}
&[~G_{\sigma\nu}r^{\nu}]_{\Sigma}=0,
\end{align}
where $r^{\nu}$ denotes a unit normal radial vector. By using the modified field equations \eqref{8}-\eqref{10} along with the Eq. \eqref{42}, we get the expression $[~\tilde{T}_{\sigma}^{\nu}r^{\sigma}]_{\Sigma}=0$, which implies $[-p_{r}-{p_{r}}^{(\Phi)}-\mathcal{P}_{r}]_{\Sigma}=0$. Therefore, we have
\begin{align}\label{43}
-p_{R}-{p_{R}}^{(\Phi)}-\mathcal{P}_{R}=0,
\end{align}
where $p_{R}\equiv p(R)$, ${p_{R}^{(\Phi)}}\equiv p^{(\Phi)}(R)$ and $\mathcal{P}_{R}\equiv \mathcal{P}(R)$. Eventually, the condition \eqref{43} may be expressed such that
\begin{align}\label{44}
\tilde{p_{R}}\equiv p_{R}-e^{-\sigma}\bigg\{\Phi'(\frac{2}{R}+\frac{\chi'}{2})
-\frac{\omega_{BD}\Phi'^{2}}{2\Phi}\bigg\}
+f^{*}\bigg\{\frac{1}{R^2}+\frac{\nu'}{R}\bigg\}+\frac{{g}'^{*}e^{-\sigma}}{R}=0.
\end{align}
The above equation means that the effective radial pressure should disappear at the surface. Note that this expression also contains a few dark source terms because of the field. The Eqs. \eqref{41} and \eqref{44} contain the necessary and sufficient conditions for matching the interior metric \eqref{5} with the exterior Schwarzschild metric \eqref{40}.

\section{Polytropic Equation of State}

This section is dedicated to examining the function of polytropic EoS in the BD theory. The use of polytropic EoS to study the stellar structure is very important as this can be used to describe a large number of different characteristics of the concerned system. In the previous sections, all observations are generic, without any specification of sources $({T_{\sigma}^{\nu}}^{(m)}, {T_{\sigma}^{\nu}}^{(\Phi)}, \vartheta_{\sigma}^{\nu})$ for the proposed system. Now, we consider $\vartheta_{\sigma}^{\nu}$ tensor as a polytrope to check its consequences on other generic sources which are highlighted in Eqs. \eqref{24}--\eqref{26}. In doing so, we consider the subsequent polytropic EoS which describes the relationship between energy density and pressure of the tensorial quantity $\vartheta_{\sigma}^{\nu}$. Thus, we have
\begin{align}\label{45}
-\vartheta_{1}^{1}=\mathcal{K}(\vartheta_{0}^{0})^\Gamma\neq-\vartheta_{2}^{2},
\end{align}
with $\Gamma=1+\frac{1}{n}$, where $n, \mathcal{K},$ and $\Gamma$ are the polytropic index, polytropic constant, and polytropic exponent, respectively. In the above expression, the temperature is implicitly described by the parameter $\mathcal{K} > 0$ that is controlled by the thermic properties of a distinct polytrope. By using Eqs. \eqref{27} and \eqref{28} in expression \eqref{45}, we get 1st order nonlinear DE for deformation factor $f^{*}$, which is provided as
\begin{align}\nonumber
f^{*}\bigg[\Phi''+\frac{1}{r^2}+\frac{2\Phi'}{r}+\frac{\omega_{BD}\Phi'^2}{2\Phi}\bigg]+{f}'^{*}
\bigg[\frac{1}{r}+\frac{\Phi'}{2}\bigg]=-\bigg(\frac{1}{\mathcal{K}}\bigg)^{\frac{1}{\Gamma}}\bigg[f^{*}\bigg
(\frac{1}{r^2}+\frac{\nu'}{r}&+\frac{2\Phi'}{r}+\frac{\nu'\Phi'}{2}-\frac{\omega_{BD}\Phi'^2}{2\Phi}
\bigg)\\\label{46}&+\frac{e^{-\sigma}{g}'^{*}}{r}
+\frac{e^{-\sigma}{g}'^{*}\Phi'}{2}\bigg]^\frac{1}{\Gamma}.
\end{align}
This result depends upon the deformation factors due to polytrope, metric potential, scalar field, and other extra curvature factors of the BD theory. Thereby, it reveals the aspects of strong gravitational interaction. To solve the system of Eqs. \eqref{27}--\eqref{29}, we establish two auxiliary conditions. These two conditions mimic constraints for pressure and density. In general, the system must be solved by adding an additional constraint to each condition. We impose a pressure mimic constraint, namely, $\vartheta_{1}^{1}\sim T_{1}^{1}$, which read as
\begin{align}\label{47}
\vartheta_{1}^{1}=\beta(\mathcal{K},\Gamma)T_{1}^{1},
\end{align}
where $\beta(\mathcal{K}, \Gamma)$ is a known dimensionless function that can be used for any polytrope. Its simplest expression is consistent with the Eq. \eqref{45} and condition $f^{*}_{r}|_{\mathcal{K}=0}=0$, written as
\begin{align}\label{48}
\beta(\mathcal{K},\Gamma)=\mathcal{K}^{\Gamma},
\end{align}
on substituting the value of Eq. \eqref{48} in the expression \eqref{47}, brings out following result
\begin{align}\label{49}
\vartheta_{1}^{1}=\mathcal{K}^\Gamma (T_{1}^{1}).
\end{align}
Now, by utilizing Eqs. \eqref{46} and \eqref{49}, we get
\begin{align}\label{50}
&f^{*}\bigg[\Phi''+\frac{1}{r^2}+\frac{2\Phi'}{r}
+\frac{\omega_{BD}\Phi'^2}{2\Phi}\bigg]+{f}'^{*}\bigg[\frac{1}{r}
+\frac{\Phi'}{2}\bigg]=-\mathcal{K}^\frac{\Gamma-1}{\Gamma}\bigg[-\frac{1}{r^2}
+e^{-\sigma}\bigg(\frac{1}{r^2}+\frac{\chi'}{r}+\frac{2 \Phi'}{r}+\frac{\chi'\Phi'}{2}-\frac{\omega_{BD}\Phi'^2}{2 \Phi}\bigg)\bigg]^{\frac{1}{\Gamma}},\\\label{51}
&{g}'^{*}=\frac{1}{(1+\frac{\Phi'r}{2})(e^{-\sigma}+f^{*})}\bigg[\bigg(\frac{1}{r}+\chi'+2 \Phi'+\frac{\chi'\Phi'r}{2}-\frac{\omega_{BD}\Phi'^2r}{2\Phi}\bigg)\bigg(\mathcal{K}^\Gamma e^{-\sigma}-f^{*}\bigg)-\frac{\mathcal{K}^\Gamma}{r}\bigg].
\end{align}
For a specified $(\chi,\sigma)$ seed solution to the modified field equations \eqref{24}--\eqref{26}, we can determine its deformation $(f^{*}, g^{*})$ which is obtained for some polytrope $(\mathcal{K}, n)$ by using Eqs. \eqref{50} and \eqref{51}. It allows us to determine the ramifications of polytrope on any non-specified fluid satisfying the Eqs.\eqref{24}-\eqref{26} regardless of its nature. Another choice, which is similarly helpful to assure a solution with physical viability, is to take the mimic-constraint for density as $\vartheta_{0}^{0}\sim T_{0}^{0}$. We have the following equation by using the same justification as provided for Eqs. \eqref{47} and \eqref{48}. As a result, we obtain
\begin{align}\label{52}
\vartheta_{0}^{0}=\mathcal{K}^\Gamma T_{0}^{0},
\end{align}
which results in the following form
\begin{align}\label{53}
&f^{*}\bigg[\Phi''+\frac{1}{r^2}+\frac{2\Phi'}{r}+\frac{\omega_{BD}\Phi'^2}{2\Phi}\bigg]
+{f}'^{*}\bigg[\frac{1}{r}+\frac{\Phi'}{2}\bigg]=-\beta \bigg[\frac{1}{r^2}-e^{-\sigma}\bigg(\frac{1}{r^2}-\frac{\sigma'}{r}+\Phi''
+\frac{2\Phi'}{r}-\frac{\Phi'\sigma'}{2}+\frac{\omega_{BD} \Phi'^{2}}{2\Phi}\bigg)\bigg],\\\nonumber
&{g}'^{*}=\frac{\mathcal{K}\beta^{\Gamma}r}{(e^{-\sigma}+f^{*})(1+\frac{\Phi'r}{2})}\bigg[\frac{1}{r^2}
-e^{-\sigma}\bigg(\frac{1}{r^2}-\frac{\sigma'}{r}+\Phi''+\frac{2\Phi'}{r}
-\frac{\Phi'\sigma'}{2}+\frac{\omega_{BD} \Phi'^{2}}{2\Phi}\bigg)\bigg]^{\Gamma}-f^{*}\bigg(\frac{1}{r^2}+\frac{\chi'}{r}
+\frac{2\Phi'}{r}+\frac{\chi'\Phi'}{2}\\\label{54}&
-\frac{\omega_{BD}\Phi'^2}{2\Phi}\bigg).
\end{align}
One can find $f^{*}(r)$ by considering both constraints presented in Eqs. \eqref{49} and \eqref{52}. The difference is that when we impose the constraint \eqref{49}, we must solve the nonlinear differential Eq. \eqref{50}. On the other hand, by imposing constraint \eqref{52}, we have to solve the linear DE \eqref{53}. In the next section, we will use the Tolman IV solution for solving the Eqs. \eqref{24}--\eqref{26}. After that, we will have the EoS \eqref{45} to examine the properties of a polytropic fluid including $\{\mathcal{K}, n\}$.
For this purpose, we examine its aspects through mimic-constraints on pressure and density stated by \eqref{49} and \eqref{52}, respectively.

\section{Polytropes supporting Perfect Fluid with Tolman IV Geometry}

Tolman \cite{tolman1939static} studied eight solutions which describe a static sphere for a perfect fluid in the framework of GR. Tolman IV and VII are two of them which are workable physical solutions. Here, we consider the familiar Tolman IV solution as a seed ($\chi, \sigma, \rho$, $p$) for perfect fluids. Consequently, we have
\begin{align}\label{55}
&e^{\chi(r)}=B^{2}\bigg(1+\frac{r^2}{A^2}\bigg);~~~~~~
e^{\sigma(r)}=\frac{1+\frac{2r^2}{A^2}}{(1-\frac{r^2}{C^2})(1
+\frac{r^2}{A^2})},\\\label{56}
&\rho(r)+\rho^{(\Phi)}(r)=\frac{3A^4+2r^2(C^2+3r^2)+A^2(3C^2+7r^2)}{C^2(A^2+2r^2)^2},\\\label{57}
&p(r)+p^{(\Phi)}(r)=\frac{C^2-A^2-3r^2}{C^2(A^2+2r^2)}.
\end{align}
In GR ($f^{*}_{r}=g^{*}_{r}=0$), the set of parameters $A, B$, and $C$ in above equations describing the geometry and physical characteristics of star structure, which is determined by matching conditions \eqref{41} and \eqref{44}, yield
\begin{align}\label{58}
A^2=\frac{R^3-3M R^2}{M},\quad\quad\quad\quad B^2=\frac{R-3M}{R},\quad\quad\quad\quad C^2=\frac{R^3}{M},
\end{align}
with the compactness $\frac{M}{R}< \frac{4}{9}$, where $M$ represents the total mass of the system. The geometric continuity is guaranteed by the formula in Eq. \eqref{58} at $R=r$ and will vary by adding $\vartheta_{\sigma}^{\nu}$. We can determine the geometric deformation by using the polytropic index $``n"$, Eq. \eqref{55} in Eq. \eqref{50}, and is expressed as
\begin{align}\label{59}
f^{*}(r)=-\frac{r^2}{3}\bigg[(\mathcal{K})^{\frac{1}{n+1}}\bigg(\frac{C^2-A^2}{A^2C^2}\bigg)^\frac{n}{n+1}\bigg\{e^{-\zeta}\bigg(\Phi'\bigg(\frac{2}{r}+\frac{\nu'}{2}\bigg)
-\frac{\omega_{BD}\Phi'^{2}}{2\Phi}\bigg)\bigg\}^{\frac{n}{n+1}}\check{H}(r)+\check{X}(r)\bigg]+\frac{c_1}{r},
\end{align}
where values for $\check{X}(r)$ and $\check{H}(r)$ are mentioned in Appendix A. It is found that the deformation due to polytrope has the correspondence with the dark source components of the BD theory within the surface of the radius. The continuity of the first fundamental form is represented in Eqs. \eqref{41} has the form 
\begin{align}\label{60}
B^2(1+\frac{R^2}{A^2})e^{g^{*}(R)}&=1-\frac{2\mathcal{M}}{R},\\\label{61}
e^{-\sigma}+f^{*}(R)&=1-\frac{2\mathcal{M}}{R},
\end{align}
where $f^{*}_{R}=f^{*}(R)$ is deformation calculated at the stellar surface and the continuity of second fundamental form in Eq. \eqref{44}
\begin{align}\label{62}
C^2&=A^2+3R^2.
\end{align}
If we consider the integration constant $c_1=0$ in Eq. \eqref{60} to have a regular solution at the origin $(r=0)$, then deformation turns out to be
\begin{align}\label{63}
f^{*}(r)+\frac{r^2}{3}\check{X}(r)=\bigg(\frac{\mathcal{K}}{3}\bigg)^{\frac{1}{n+1}}r^2
\bigg\{\frac{R^2}{A^2(A^2+3R^2)}\bigg\}^{\frac{n}{n+1}}\bigg\{e^{-\zeta}\bigg[\Phi'\bigg(\frac{2}{r}+\frac{\nu'}{2}\bigg)
-\frac{\omega_{BD}\Phi'^{2}}{2\Phi}\bigg]\bigg\}^{\frac{n}{n+1}}~\check{H}(r).
\end{align}
For the Schwarzschild mass, we get the following expression by applying the condition \eqref{61}, so that
\begin{align}\label{64}
\frac{2~m(R)}{R}-f^{*}_{R}=\frac{2~\mathcal{M}}{R},
\end{align}
where $m(R)$ is used in the expression \eqref{21}. Finally, by the combination of the above expression and junction condition \eqref{60}, we obtain the following result
\begin{align}\label{65}
B^2(1+\frac{R^2}{A^2})e^{g^{*}(R)}=\frac{A^2+R^2}{A^2+3R^2}+f^{*}(R).
\end{align}
This shows that the constants in expressions \eqref{58} are the functions of $(K,n)$ which are polytropic variables. To match the interior metric \eqref{5} to an exterior metric in Eq. \eqref{40} requires that Eqs. \eqref{62}, \eqref{64} and \eqref{65} must be satisfied. By using Eqs.\eqref{16}, \eqref{45}, \eqref{58} and \eqref{62}, we obtain the total pressure in following expression as
\begin{align*}
\tilde{p_{r}}=\frac{3(R^2-r^2)(1+\mathcal{K}^\Gamma)+\mathcal{K}^\Gamma e^{-\zeta}(A^2+3R^2)(A^2+2r^2)\bigg(\Phi'\bigg(\frac{2}{r}+\frac{\nu'}{2}\bigg)
-\frac{\omega_{BD}\Phi'^{2}}{2\Phi}\bigg)}{(A^2+3R^2)(A^2+2r^2)}.
\end{align*}
The effective density in Eq.~\eqref{15} and effective tangential pressure which is obtained by Eq.~\eqref{17}, have analytical formulations in terms of $(\mathcal{K},n)$, but they are too massive to present. As we observe, our solution eliminates the need for perturbative analysis, but they play an essential role in understanding the behavior of incredibly dense stellar structures. For the feasible characteristics of a self-gravitating system, state determinants must be finite, positive, maximum at center and monotonically decreasing toward the stellar structure.
%The energy density with the help of Eq.(57) is given as
%\begin{align}\label{67}
%\varepsilon=A(3F_{n}+rF'_{n})+\frac{1}{3}(3F_{\Phi}+rF'_{\Phi})+(AF_{n}+\frac{F_{\Phi}}{3})(\Phi''r^2+\frac{\omega_{BD}\Phi'^2r^2}{2\Phi})
%+r\Phi'A(3F_{n}+\frac{rF_{n}'}{2})+\frac{r\Phi'}{3}(3F_{\Phi}+\frac{rF_{\Phi}'}{2})
%\end{align}

%The comportment of these matter variables against the radial coordinate should be positive and decrease monotonically towards the stellar surface. Figs. 2 and 3 reveal that the state determinants are maximum at the center which exhibits that the core of compact configuration is highly concentrated and decrease away from it for the selected values of the parameters , , and .
The model under consideration is a scalar-tensor model involving two distinct fluids. The implications of the analytical solutions are perhaps limited for this particular setup, and one can resort to numerical techniques that allow for the exploration of the model's behavior and predictions through computational simulations.

%%%%%%%%%%%%%%%%%%%%%%%%%%%%%%%%%%%%%%%%%%%%%%%%%%%%%%%%%%
\begin{figure}[t]\centering
\epsfig{file=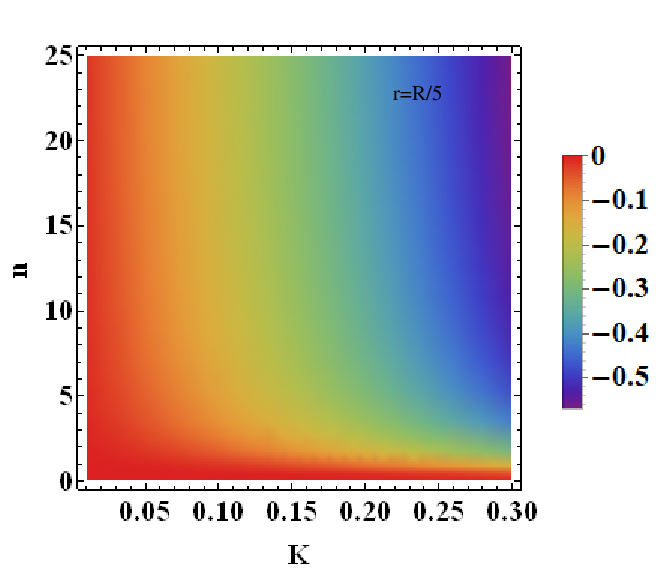,width=0.3\linewidth}
\epsfig{file=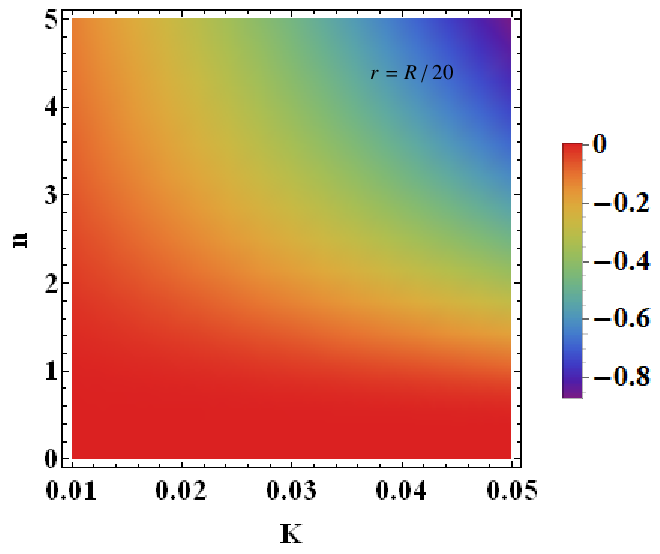,width=0.3\linewidth}
\caption{Radial pressure $\tilde{\emph{p}}^{(eff)}_{r}(\mathcal{K},n)$ for $r=\frac{R}{5}$ (the left panel) and for $r=\frac{R}{20}$ (the right panel) on $\mathcal{K}$-$n$ plane.} \label{1}
\end{figure}
\begin{figure}[t]\centering
\epsfig{file=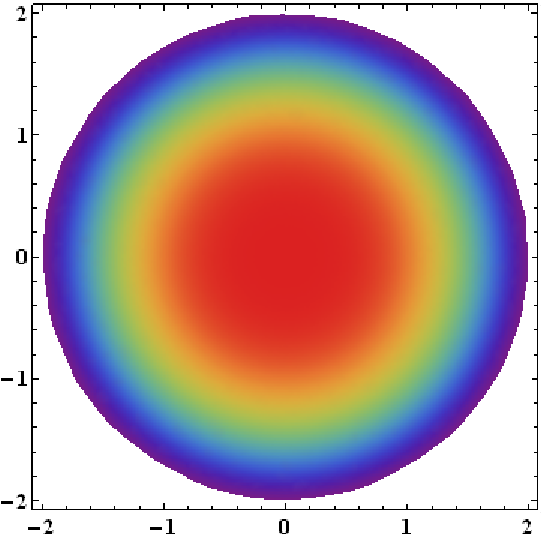,width=0.3\linewidth}
\epsfig{file=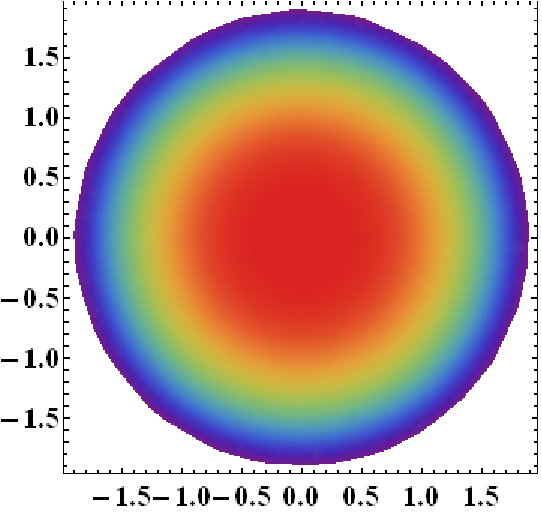,width=0.3\linewidth}
\caption{Radial pressure $\tilde{\emph{p}}_{r}(r)$ for $n= 0.5$ (the left panel)and $n= 0.8$ (the right panel). From these figures, the impact of the polytrope on the ideal fluid ($\mathcal{K}=0$) is observed. The horizontal axis indicates the radial coordinate, while the vertical axis shows the radial pressure.} \label{2}
\end{figure}
\begin{figure}[t]\centering
\epsfig{file=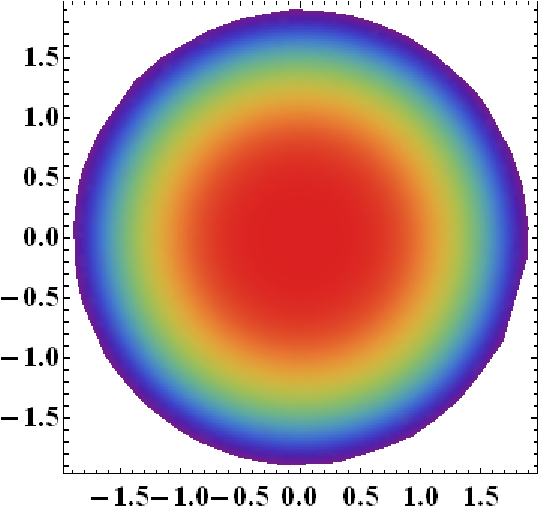,width=0.3\linewidth}
\epsfig{file=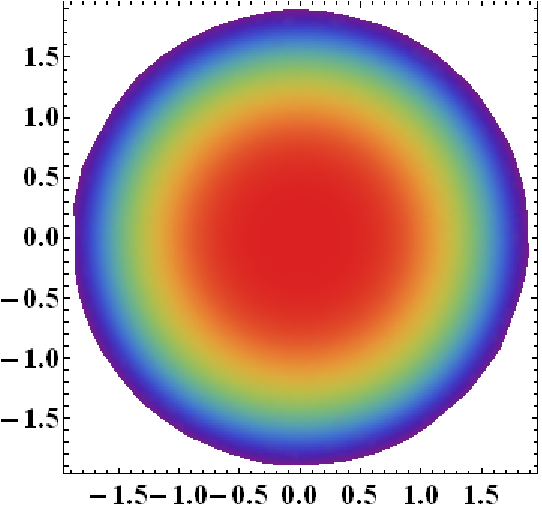,width=0.3\linewidth}
\caption{Radial pressure $\tilde{\emph{p}}_{r}(r)$ for $\mathcal{K}=0.01$ and $n= 0.5$ (the left panel), $n= 0.8$ (the right panel). The horizontal and vertical axes are the same as in Fig. \textbf{\ref{2}}.} \label{3}
\end{figure}
\begin{figure}[t]\centering
\epsfig{file=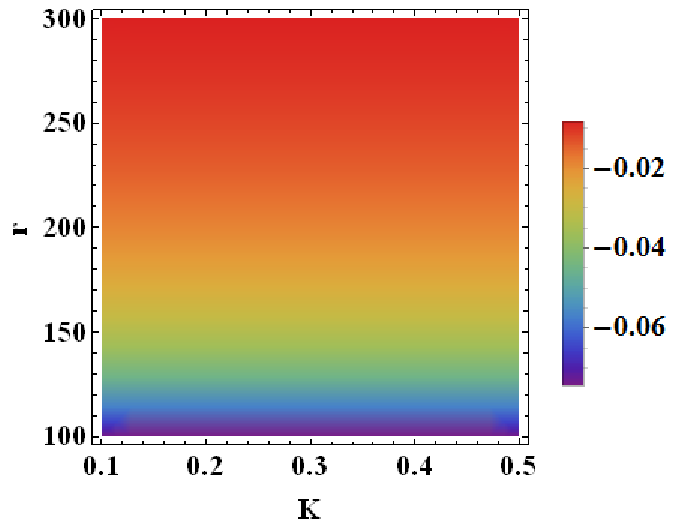,width=0.3\linewidth}
\epsfig{file=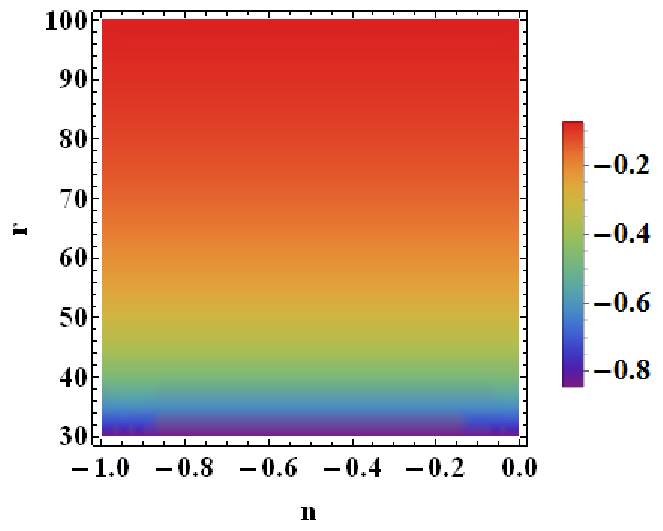,width=0.3\linewidth}
\caption{Pressures $\tilde{\emph{p}}_{r}(\mathcal{K},r)$ (the left panel) and $\tilde{\emph{p}}_{r}(n,r)$ (the right panel).} \label{4}
\end{figure}
\begin{figure}[t]\centering
\epsfig{file=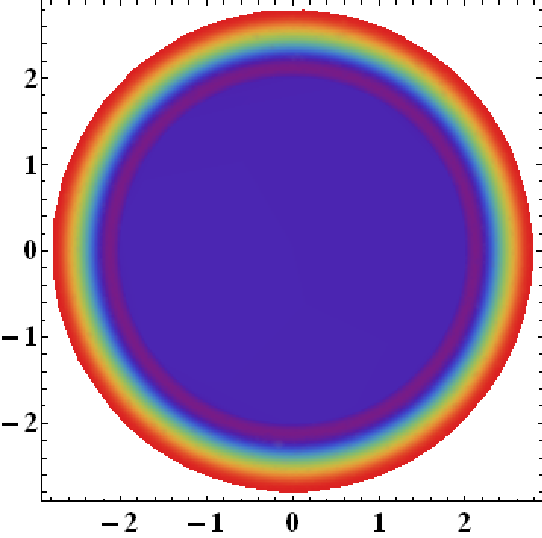,width=0.3\linewidth}
\epsfig{file=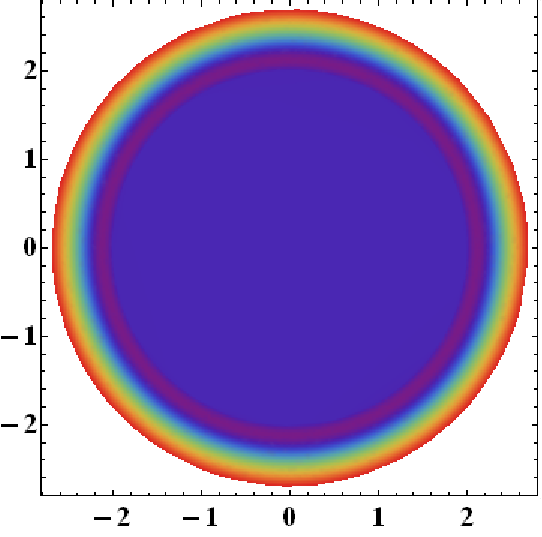,width=0.3\linewidth}
\caption{Anisotropy (the left panel) and fluid energy exchange (the right panel) for $\mathcal{K}=0.01$ and $n=0.5$. The horizontal axis shows the radial coordinate, while the vertical axis indicates the anisotropy.} \label{5}
\end{figure}
%%%%%%%%%%%%%%%%%%%%%%%%%%%%%%%%%%%%%%%%%%%%%%%%%%%%%%%%%%

We depict the stellar solutions for the model considered in this manuscript. Figure \textbf{\ref{1}} represents the effects of radial pressure and the BD theory corrections for relativistic gravitational spheres. The plots of $\tilde{p_{r}}(n,r)$ describe the effective matter variables, which are proportional to $\mathcal{K}$ and have an inverse relation with $n$.
%describe adequate trends, including maximum at the center and finite throughout.
A similar kind of the behavior is observed in Figs. \textbf{\ref{2}} and \textbf{\ref{3}}, which show the contribution of polytropic scalar field radial pressure for different values of the polytropic constant. However, Fig. \textbf{\ref{4}} represents the $\tilde{p_{r}}(\mathcal{K},r)$ and $\tilde{p_{r}}(n,r)$. It is clearly seen from Fig. \textbf{\ref{4}} that when $\mathcal{K}$ increases, the contribution of the polytropic scalar field radial pressure increases by keeping $n$ constant. Finally, the interaction between fluids that produce anisotropy and a positive energy gradient with respect to $r$ inside the stellar distribution is shown in Fig. \textbf{\ref{5}}.

\section{Conclusions}

The compact stars are highly dense because of their small size and reveal some enigmatic characteristics of the universe. Modified gravity theories can lead to further understanding of their evolution. In this paper, we explored the self-gravitating systems which demand anisotropic solutions of the modified field equations. For this purpose, the gravitational decoupling technique has been used by taking the line element for the static system in the context of the BD theory. This theory provided a scalar field that describes how the universe evolved. Also, it has a possible function that matches observational data with parameters determining the inflationary era.

In the present paper, we have explored the modified field equations in expression \eqref{8}--\eqref{10} to examine the motion of our system, where a new gravitational source and field terms are added to the stress-energy tensor of an imperfect fluid. In this regard, we have successfully decoupled the modified field equations by distorted temporal and radial metric potentials by the GD method. We have also discussed the MGD technique by taking only the deformation of the radial component of the metric potentials. The Bianchi Identities examined for both sources with dark constituents of the BD theory, as a result, a specific relationship between them is shown in expressions \eqref{34} and \eqref{35}. The analysis of the combined effective solution constructed by the two fluids together with dark source terms of the field yields energy gradients that increase in the radial direction. By utilizing the Schwarzschild space-time and the GD interior line element and by applying matching conditions, we have determined the surface at which the effective pressure along $r$ becomes zero.

Furthermore, we have chosen the polytropic fluid specified by $\mathcal{K}$ and $n$ to write the system in another form with polytropic EoS, which describes a large variety of physical phenomena in compact objects. As a straightforward application, we have analyzed the situation where the coexistence of a polytrope with a perfect fluid is defined by the variables $(\mathcal{K}, \Gamma)$. To complete the system, we specifically used the famous Tolman IV perfect fluid solution as a seed which has the limit $K\rightarrow 0$, when the consequences of polytropic fluid disappear.

In addition, we have found that the perfect fluid receives the energy transfer from the polytrope. The results obtained meet all requirements for physical acceptability including being regular at the origin and having positive pressure also a monotonically decreasing behavior of pressure, which satisfies the strong energy conditions. It is interesting to mention that for the BD vacuum solutions where $\Phi$ = constant and $\omega_{BD} \rightarrow \infty$, all our results are compatible with those that exist in GR \cite{ovalle2022energy}.

The gravitational interaction's behavior at extreme conditions, such as in the early phases of the cosmos or in regions of high curvature, is referred to as the GD.
Since the BD theory suggests a violation of the equivalence principle, we may develop studies to investigate the equivalence principle in the solar system by investigating GD in the BD theory, such as precise observations of planet motion or light deflecting under the influence of massive bodies.
It can also reveal insights into fundamental physics, such as the behavior of scalar fields, which other modified theories cannot examine.
Furthermore, knowing how gravity decouples in the BD theory aids in expanding our comprehension of the universe's growth, the development of celestial objects, and the behavior of astrophysical objects and has implications for gravitational wave physics. It can also illuminate situations such as inflation and cosmic microwave background radiation.
Anisotropy can cause deviations from traditional general relativity predictions near the surface of an object or in its core. Studying these aberrations can provide explanations for the behavior of matter under severe gravitational circumstances, allowing gravitational theories to be tested and revealing information about the physics of black holes. Finally, it is emphasized that the analysis performed in this work can be studied in any other gravitational theory with modification in the Hilbert action, where we can examine the possible exchange of energy. It is still an open question to study the stability for the case that there are radiations in the interior or the seed source is dissipative and viscous.

There are indeed many extensions and modifications of scalar-tensor theories, on which researchers are working including those involving mass terms, self-interacting potentials, nonlinear kinetic terms, more general non minimal coupling, and Horndeski theories. However, the complex extensions of scalar-tensor theories can be computationally intensive to study, especially in situations where analytical solutions are not readily available. It is worth mentioning that gravitational decoupling involves complex interactions between gravity and scalar fields so restricting our analysis to a specific scalar-tensor theory helps to maintain the theoretical relevance of the study.
The additional contribution ($\vartheta_{\sigma}^{\nu}$) to our proposed system makes the system highly complex. The key feature of the BD theory is that it provides a straightforward expanding solution for the scalar field consistent with observations of the solar system. In this regard, the particular scalar-tensor theory can be advantageous without introducing additional complexities from extensions.
It is significant to compare the behavior of gravitational decoupling in different scalar-tensor theories to understand how various theoretical choices impact the decoupling process, however, a comprehensive examination of gravitational decoupling within the broader context of scalar-tensor theories is unexplored.

%\vspace{0.5cm}

%\noindent {\bf Acknowledgement:}
\section*{Acknowledgement}
The work of KB was supported in part by the JSPS KAKENHI Grant Number JP21K03547.

%\vspace{0.5cm}
%\noindent {\bf Data Availability Statement:}
%%%%%
%%%%%
%\section*{Data Availability Statement}
%This manuscript has no associated data or the data will not be deposited.
%%%%%
%%%%%

%\vspace{0.5cm}
%\noindent {\bf Declaration of Competing Interest:}
%%%%%
%%%%%
%\section*{Declaration of Competing Interest}
%The authors declare that they have no known competing financial interests or pe%rsonal relationships that could have appeared to influence the work reported in% this paper.
%%%%%
%%%%%

\section*{Appendix A}

In this Appendix, we describe several detailed expressions discussed in Sec. III.
The representations of ${T_{0}^{0}}^{(eff)}$, ${T_{1}^{1}}^{(eff)}$, and ${T_{2}^{2}}^{(eff)}$ in the right-hand side of Eq. \eqref{34} are given as
\begin{align*}
{T_{0}^{0}}^{(eff)}&=T_{0}^{0}+e^{-\sigma}\bigg(\Phi''+\frac{2\Phi'}{r}
+\frac{\omega_{BD}\Phi'^{2}}{2\Phi}\bigg)+\frac{(e^{-\sigma})'\Phi'}{2},\\
{T_{1}^{1}}^{(eff)}&=T_{1}^{1}+e^{-\sigma}(\frac{2\Phi'}{r}+\frac{\chi'\Phi'}{2}
-\frac{\omega_{BD}\Phi'^{2}}{2\Phi}\bigg),\\
{T_{2}^{2}}^{(eff)}&=T_{2}^{2}+e^{-\sigma}\bigg(\Phi''+\frac{\Phi'}{r}
+\frac{\Phi'\nu'}{2}+\frac{\omega_{BD}\Phi'^{2}}{2\Phi}\bigg)
+\frac{(e^{-\sigma})'\Phi'}{2}.
\end{align*}
The expression for $\check{X}(r)$ and $\check{H}(r)$ which appeared in the right hand side of Eq.(\ref{59}) are given as follows
\begin{align*}
&\check{X}(r)=\frac{r^{3}}{3}\bigg(f^{*}\Phi''+\frac{{f}'^{*}\Phi'}{2}+\frac{2\Phi'f^{*}}{r}+
\frac{\omega_{BD}\Phi'^{2}f^{*}}{2\Phi}\bigg)
-\int\frac{r^{3}}{3}\frac{d}{dr}\bigg(f^{*}\Phi''+\frac{{f}'^{*}\Phi'}{2}
+\frac{2\Phi'f^{*}}{r}+\frac{\omega_{BD}\Phi'^{2}f^{*}}{2\Phi}\bigg)dr,\\\nonumber
&\check{H}(r)=\frac{1}{p_{r}^{(\Phi)}}\bigg[\bigg(S_{1}+\frac{A^2C^2}{(C^2-A^2)}\bigg)^\frac{n}{n+1}-\frac{1}{r^3}\int \bigg\{r^3\frac{d}{dr}\bigg(S_{1}+e^{-\zeta}\bigg(\Phi'\bigg(\frac{2}{r}+\frac{\nu'}{2}\bigg)
-\frac{\omega_{BD}\Phi'^{2}}{2\Phi}\bigg)
A^2C^2(C^2-A^2)^{-1}\bigg)^{\frac{n}{n+1}}\bigg\}dr\bigg],
\end{align*}
with
\begin{align*}
S_{1}=1-\frac{r^2(A^2+2C^2)}{(A^2+2r^2)(C^2-A^2)}.
\end{align*}


\vspace{0.5cm}

\begin{thebibliography}{77}
\expandafter\ifx\csname natexlab\endcsname\relax\def\natexlab#1{#1}\fi
\expandafter\ifx\csname bibnamefont\endcsname\relax
  \def\bibnamefont#1{#1}\fi
\expandafter\ifx\csname bibfnamefont\endcsname\relax
  \def\bibfnamefont#1{#1}\fi
\expandafter\ifx\csname citenamefont\endcsname\relax
  \def\citenamefont#1{#1}\fi
\expandafter\ifx\csname url\endcsname\relax
  \def\url#1{\texttt{#1}}\fi
\expandafter\ifx\csname urlprefix\endcsname\relax\def\urlprefix{URL }\fi
\providecommand{\bibinfo}[2]{#2}
\providecommand{\eprint}[2][]{\url{#2}}

\bibitem[{\citenamefont{Schwarzschild}(1916)}]{schwarzschild1916sitzungsber}
\bibinfo{author}{\bibfnamefont{K.}~\bibnamefont{Schwarzschild}},
  \bibinfo{journal}{Preuss. Akad. Wiss. Berlin Math. Phys.}
  \textbf{\bibinfo{volume}{189}}, \bibinfo{pages}{1916} (\bibinfo{year}{1916}).

\bibitem[{\citenamefont{Tolman}(1939)}]{tolman1939static}
\bibinfo{author}{\bibfnamefont{R.~C.} \bibnamefont{Tolman}},
  \bibinfo{journal}{Phys. Rev.} \textbf{\bibinfo{volume}{55}},
  \bibinfo{pages}{364} (\bibinfo{year}{1939}).

\bibitem[{\citenamefont{Perlmutter et~al.}(1997)}]{perlmutter1997measurements}
\bibinfo{author}{\bibnamefont{Perlmutter}} \bibnamefont{et~al.},
  \bibinfo{journal}{Astrophys. J.} \textbf{\bibinfo{volume}{483}},
  \bibinfo{pages}{565} (\bibinfo{year}{1997}).

\bibitem[{\citenamefont{Perlmutter et~al.}(1998)}]{perlmutter1998discovery}
\bibinfo{author}{\bibnamefont{Perlmutter}} \bibnamefont{et~al.},
  \bibinfo{journal}{Nature} \textbf{\bibinfo{volume}{391}}, \bibinfo{pages}{51}
  (\bibinfo{year}{1998}).

\bibitem[{\citenamefont{Filippenko and Riess}(1998)}]{filippenko1998results}
\bibinfo{author}{\bibfnamefont{A.~V.} \bibnamefont{Filippenko}}
  \bibnamefont{and} \bibinfo{author}{\bibfnamefont{A.~G.} \bibnamefont{Riess}},
  \bibinfo{journal}{Phys. Rep.} \textbf{\bibinfo{volume}{307}},
  \bibinfo{pages}{31} (\bibinfo{year}{1998}).

\bibitem[{\citenamefont{Caldwell et~al.}(2003)\citenamefont{Caldwell,
  Kamionkowski, and Weinberg}}]{caldwell2003phantom}
\bibinfo{author}{\bibfnamefont{R.~R.} \bibnamefont{Caldwell}},
  \bibinfo{author}{\bibfnamefont{M.}~\bibnamefont{Kamionkowski}},
  \bibnamefont{and} \bibinfo{author}{\bibfnamefont{N.~N.}
  \bibnamefont{Weinberg}}, \bibinfo{journal}{Phys. Rev. Lett.}
  \textbf{\bibinfo{volume}{91}}, \bibinfo{pages}{071301}
  (\bibinfo{year}{2003}).

\bibitem[{\citenamefont{Tegmark et~al.}(2004)}]{tegmark2004cosmological}
\bibinfo{author}{\bibnamefont{Tegmark}} \bibnamefont{et~al.},
  \bibinfo{journal}{Phys. Rev. D} \textbf{\bibinfo{volume}{69}},
  \bibinfo{pages}{103501} (\bibinfo{year}{2004}).

\bibitem[{\citenamefont{Land and Magueijo}(2005)}]{land2005examination}
\bibinfo{author}{\bibfnamefont{K.}~\bibnamefont{Land}} \bibnamefont{and}
  \bibinfo{author}{\bibfnamefont{J.}~\bibnamefont{Magueijo}},
  \bibinfo{journal}{Phys. Rev. Lett.} \textbf{\bibinfo{volume}{95}},
  \bibinfo{pages}{071301} (\bibinfo{year}{2005}).

\bibitem[{\citenamefont{Sherwin et~al.}(2011)}]{sherwin2011evidence}
\bibinfo{author}{\bibnamefont{Sherwin}} \bibnamefont{et~al.},
  \bibinfo{journal}{Phys. Rev. Lett.} \textbf{\bibinfo{volume}{107}},
  \bibinfo{pages}{021302} (\bibinfo{year}{2011}).

\bibitem[{\citenamefont{Stuchl\'{i}k et~al.}(2016)\citenamefont{Stuchl\'{i}k,
  Hled\'{i}k, and Novotn\'{y}}}]{stuchlik2016general}
\bibinfo{author}{\bibfnamefont{Z.}~\bibnamefont{Stuchl\'{i}k}},
  \bibinfo{author}{\bibfnamefont{S.}~\bibnamefont{Hled\'{i}k}},
  \bibnamefont{and}
  \bibinfo{author}{\bibfnamefont{J.}~\bibnamefont{Novotn\'{y}}},
  \bibinfo{journal}{Phys. Rev. D} \textbf{\bibinfo{volume}{94}},
  \bibinfo{pages}{103513} (\bibinfo{year}{2016}).

\bibitem[{\citenamefont{Novotn\'{y} et~al.}(2017)\citenamefont{Novotn\'{y},
  Hlad\'{i}k, and Stuchl{\'\i}k}}]{novotny2017polytropic}
\bibinfo{author}{\bibfnamefont{J.}~\bibnamefont{Novotn\'{y}}},
  \bibinfo{author}{\bibfnamefont{J.}~\bibnamefont{Hlad\'{i}k}},
  \bibnamefont{and}
  \bibinfo{author}{\bibfnamefont{Z.}~\bibnamefont{Stuchl{\'\i}k}},
  \bibinfo{journal}{Phys. Rev. D} \textbf{\bibinfo{volume}{95}},
  \bibinfo{pages}{043009} (\bibinfo{year}{2017}).

\bibitem[{\citenamefont{Stuchl\'{i}k et~al.}(2017)\citenamefont{Stuchl\'{i}k,
  Schee, Toshmatov, Hlad\'{i}k, and Novotn\'{y}}}]{stuchlik2017gravitational}
\bibinfo{author}{\bibfnamefont{Z.}~\bibnamefont{Stuchl\'{i}k}},
  \bibinfo{author}{\bibfnamefont{J.}~\bibnamefont{Schee}},
  \bibinfo{author}{\bibfnamefont{B.}~\bibnamefont{Toshmatov}},
  \bibinfo{author}{\bibfnamefont{J.}~\bibnamefont{Hlad\'{i}k}},
  \bibnamefont{and}
  \bibinfo{author}{\bibfnamefont{J.}~\bibnamefont{Novotn\'{y}}},
  \bibinfo{journal}{J. Cosmol. Astropart. Phys.}
  \textbf{\bibinfo{volume}{2017}}, \bibinfo{pages}{056} (\bibinfo{year}{2017}).

\bibitem[{\citenamefont{Hod}(2018{\natexlab{a}})}]{hod2018lower}
\bibinfo{author}{\bibfnamefont{S.}~\bibnamefont{Hod}}, \bibinfo{journal}{Phys.
  Rev. D} \textbf{\bibinfo{volume}{97}}, \bibinfo{pages}{084018}
  (\bibinfo{year}{2018}{\natexlab{a}}).

\bibitem[{\citenamefont{Hod}(2018{\natexlab{b}})}]{hod2018analytic}
\bibinfo{author}{\bibfnamefont{S.}~\bibnamefont{Hod}}, \bibinfo{journal}{Eur.
  Phys. J. C} \textbf{\bibinfo{volume}{78}}, \bibinfo{pages}{417}
  (\bibinfo{year}{2018}{\natexlab{b}}).

\bibitem[{\citenamefont{Lema{\^\i}tre}(1933)}]{lemaitre1933univers}
\bibinfo{author}{\bibfnamefont{G.}~\bibnamefont{Lema{\^\i}tre}}, in
  \emph{\bibinfo{booktitle}{Annales de la Soci{\'e}t{\'e} scientifique de
  Bruxelles}} (\bibinfo{year}{1933}), vol.~\bibinfo{volume}{53},
  p.~\bibinfo{pages}{51}.

\bibitem[{\citenamefont{Sokolov}(1980)}]{sokolov1980phase}
\bibinfo{author}{\bibfnamefont{A.}~\bibnamefont{Sokolov}},
  \bibinfo{journal}{Sov. Phys. JETP} \textbf{\bibinfo{volume}{52}},
  \bibinfo{pages}{575} (\bibinfo{year}{1980}).

\bibitem[{\citenamefont{Ruderman}(1972)}]{ruderman1972pulsars}
\bibinfo{author}{\bibfnamefont{M.}~\bibnamefont{Ruderman}},
  \bibinfo{journal}{Annu. Rev. Astron. Astrophys.}
  \textbf{\bibinfo{volume}{10}}, \bibinfo{pages}{427} (\bibinfo{year}{1972}).

\bibitem[{\citenamefont{Bowers and Liang}(1974)}]{bowers1974anisotropic}
\bibinfo{author}{\bibfnamefont{R.~L.} \bibnamefont{Bowers}} \bibnamefont{and}
  \bibinfo{author}{\bibfnamefont{E.~P.~T.} \bibnamefont{Liang}},
  \bibinfo{journal}{Astrophys. J.} \textbf{\bibinfo{volume}{188}},
  \bibinfo{pages}{657} (\bibinfo{year}{1974}).

\bibitem[{\citenamefont{Maurya and Maharaj}(2017)}]{maurya2017anisotropic}
\bibinfo{author}{\bibfnamefont{S.~K.} \bibnamefont{Maurya}} \bibnamefont{and}
  \bibinfo{author}{\bibfnamefont{S.~D.} \bibnamefont{Maharaj}},
  \bibinfo{journal}{Eur. Phys. J. C} \textbf{\bibinfo{volume}{77}},
  \bibinfo{pages}{328} (\bibinfo{year}{2017}).

\bibitem[{\citenamefont{Matondo et~al.}(2018)\citenamefont{Matondo, Maharaj,
  and Ray}}]{matondo2018relativistic}
\bibinfo{author}{\bibfnamefont{D.~K.} \bibnamefont{Matondo}},
  \bibinfo{author}{\bibfnamefont{S.~D.} \bibnamefont{Maharaj}},
  \bibnamefont{and} \bibinfo{author}{\bibfnamefont{S.}~\bibnamefont{Ray}},
  \bibinfo{journal}{Eur. Phys. J. C} \textbf{\bibinfo{volume}{78}},
  \bibinfo{pages}{473} (\bibinfo{year}{2018}).

\bibitem[{\citenamefont{Ovalle}(2017)}]{ovalle2017decoupling}
\bibinfo{author}{\bibfnamefont{J.}~\bibnamefont{Ovalle}},
  \bibinfo{journal}{Phys. Rev. D} \textbf{\bibinfo{volume}{95}},
  \bibinfo{pages}{104019} (\bibinfo{year}{2017}).

\bibitem[{\citenamefont{Ovalle et~al.}(2018{\natexlab{a}})\citenamefont{Ovalle,
  Casadio, da~Rocha, and Sotomayor}}]{ovalle2018anisotropic}
\bibinfo{author}{\bibfnamefont{J.}~\bibnamefont{Ovalle}},
  \bibinfo{author}{\bibfnamefont{R.}~\bibnamefont{Casadio}},
  \bibinfo{author}{\bibfnamefont{R.}~\bibnamefont{da~Rocha}}, \bibnamefont{and}
  \bibinfo{author}{\bibfnamefont{A.}~\bibnamefont{Sotomayor}},
  \bibinfo{journal}{Eur. Phys. J. C} \textbf{\bibinfo{volume}{78}},
  \bibinfo{pages}{122} (\bibinfo{year}{2018}{\natexlab{a}}).

\bibitem[{\citenamefont{Buchdahl}(1959)}]{buchdahl1959general}
\bibinfo{author}{\bibfnamefont{H.~A.} \bibnamefont{Buchdahl}},
  \bibinfo{journal}{Phys. Rev.} \textbf{\bibinfo{volume}{116}},
  \bibinfo{pages}{1027} (\bibinfo{year}{1959}).

\bibitem[{\citenamefont{Randall and
  Sundrum}(1999{\natexlab{a}})}]{randall1999large}
\bibinfo{author}{\bibfnamefont{L.}~\bibnamefont{Randall}} \bibnamefont{and}
  \bibinfo{author}{\bibfnamefont{R.}~\bibnamefont{Sundrum}},
  \bibinfo{journal}{Phys. Rev. Lett.} \textbf{\bibinfo{volume}{83}},
  \bibinfo{pages}{3370} (\bibinfo{year}{1999}{\natexlab{a}}).

\bibitem[{\citenamefont{Randall and
  Sundrum}(1999{\natexlab{b}})}]{randall1999alternative}
\bibinfo{author}{\bibfnamefont{L.}~\bibnamefont{Randall}} \bibnamefont{and}
  \bibinfo{author}{\bibfnamefont{R.}~\bibnamefont{Sundrum}},
  \bibinfo{journal}{Phys. Rev. Lett.} \textbf{\bibinfo{volume}{83}},
  \bibinfo{pages}{4690} (\bibinfo{year}{1999}{\natexlab{b}}).

\bibitem[{\citenamefont{Ovalle}(2008)}]{ovalle2008searching}
\bibinfo{author}{\bibfnamefont{J.}~\bibnamefont{Ovalle}},
  \bibinfo{journal}{Mod. Phys. Lett. A} \textbf{\bibinfo{volume}{23}},
  \bibinfo{pages}{3247} (\bibinfo{year}{2008}).

\bibitem[{\citenamefont{Ovalle and Linares}(2013)}]{ovalle2013tolman}
\bibinfo{author}{\bibfnamefont{J.}~\bibnamefont{Ovalle}} \bibnamefont{and}
  \bibinfo{author}{\bibfnamefont{F.}~\bibnamefont{Linares}},
  \bibinfo{journal}{Phys. Rev. D} \textbf{\bibinfo{volume}{88}},
  \bibinfo{pages}{104026} (\bibinfo{year}{2013}).

\bibitem[{\citenamefont{Ovalle et~al.}(2013)\citenamefont{Ovalle, Linares,
  Pasqua, and Sotomayor}}]{ovalle2013role}
\bibinfo{author}{\bibfnamefont{J.}~\bibnamefont{Ovalle}},
  \bibinfo{author}{\bibfnamefont{F.}~\bibnamefont{Linares}},
  \bibinfo{author}{\bibfnamefont{A.}~\bibnamefont{Pasqua}}, \bibnamefont{and}
  \bibinfo{author}{\bibfnamefont{A.}~\bibnamefont{Sotomayor}},
  \bibinfo{journal}{Class. Quant. Grav.} \textbf{\bibinfo{volume}{30}},
  \bibinfo{pages}{175019} (\bibinfo{year}{2013}).

\bibitem[{\citenamefont{Casadio et~al.}(2014)\citenamefont{Casadio, Ovalle, and
  Da~Rocha}}]{casadio2014black}
\bibinfo{author}{\bibfnamefont{R.}~\bibnamefont{Casadio}},
  \bibinfo{author}{\bibfnamefont{J.}~\bibnamefont{Ovalle}}, \bibnamefont{and}
  \bibinfo{author}{\bibfnamefont{R.}~\bibnamefont{Da~Rocha}},
  \bibinfo{journal}{Class. Quant. Grav.} \textbf{\bibinfo{volume}{31}},
  \bibinfo{pages}{045016} (\bibinfo{year}{2014}).

\bibitem[{\citenamefont{Casadio
  et~al.}(2015{\natexlab{a}})\citenamefont{Casadio, Ovalle, and
  Da~Rocha}}]{casadio2015classical}
\bibinfo{author}{\bibfnamefont{R.}~\bibnamefont{Casadio}},
  \bibinfo{author}{\bibfnamefont{J.}~\bibnamefont{Ovalle}}, \bibnamefont{and}
  \bibinfo{author}{\bibfnamefont{R.}~\bibnamefont{Da~Rocha}},
  \bibinfo{journal}{Europhys. Lett.} \textbf{\bibinfo{volume}{110}},
  \bibinfo{pages}{40003} (\bibinfo{year}{2015}{\natexlab{a}}).

\bibitem[{\citenamefont{Casadio
  et~al.}(2015{\natexlab{b}})\citenamefont{Casadio, Ovalle, and
  Da~Rocha}}]{casadio2015minimal}
\bibinfo{author}{\bibfnamefont{R.}~\bibnamefont{Casadio}},
  \bibinfo{author}{\bibfnamefont{J.}~\bibnamefont{Ovalle}}, \bibnamefont{and}
  \bibinfo{author}{\bibfnamefont{R.}~\bibnamefont{Da~Rocha}},
  \bibinfo{journal}{Class. Quant. Grav.} \textbf{\bibinfo{volume}{32}},
  \bibinfo{pages}{215020} (\bibinfo{year}{2015}{\natexlab{b}}).

\bibitem[{\citenamefont{Ovalle and Sotomayor}(2018)}]{ovalle2018simple}
\bibinfo{author}{\bibfnamefont{J.}~\bibnamefont{Ovalle}} \bibnamefont{and}
  \bibinfo{author}{\bibfnamefont{A.}~\bibnamefont{Sotomayor}},
  \bibinfo{journal}{Eur. Phys. J. Plus} \textbf{\bibinfo{volume}{133}},
  \bibinfo{pages}{428} (\bibinfo{year}{2018}).

\bibitem[{\citenamefont{Ovalle et~al.}(2018{\natexlab{b}})\citenamefont{Ovalle,
  Casadio, Rocha, Sotomayor, and Stuchl{\'\i}k}}]{ovalle2018black}
\bibinfo{author}{\bibfnamefont{J.}~\bibnamefont{Ovalle}},
  \bibinfo{author}{\bibfnamefont{R.}~\bibnamefont{Casadio}},
  \bibinfo{author}{\bibfnamefont{R.~d.} \bibnamefont{Rocha}},
  \bibinfo{author}{\bibfnamefont{A.}~\bibnamefont{Sotomayor}},
  \bibnamefont{and}
  \bibinfo{author}{\bibfnamefont{Z.}~\bibnamefont{Stuchl{\'\i}k}},
  \bibinfo{journal}{Eur. Phys. J. C} \textbf{\bibinfo{volume}{78}},
  \bibinfo{pages}{960} (\bibinfo{year}{2018}{\natexlab{b}}).

\bibitem[{\citenamefont{Contreras and
  Bargue\~{n}o}(2018)}]{contreras2018minimal}
\bibinfo{author}{\bibfnamefont{E.}~\bibnamefont{Contreras}} \bibnamefont{and}
  \bibinfo{author}{\bibfnamefont{P.}~\bibnamefont{Bargue\~{n}o}},
  \bibinfo{journal}{Eur. Phys. J. C} \textbf{\bibinfo{volume}{78}},
  \bibinfo{pages}{558} (\bibinfo{year}{2018}).

\bibitem[{\citenamefont{Contreras}(2019)}]{contreras2019gravitational}
\bibinfo{author}{\bibfnamefont{E.}~\bibnamefont{Contreras}},
  \bibinfo{journal}{Class. Quant. Grav.} \textbf{\bibinfo{volume}{36}},
  \bibinfo{pages}{095004} (\bibinfo{year}{2019}).

\bibitem[{\citenamefont{Rinc{\'o}n et~al.}(2020)}]{rincon2020anisotropic}
\bibinfo{author}{\bibnamefont{Rinc{\'o}n}} \bibnamefont{et~al.},
  \bibinfo{journal}{Eur. Phys. J. C} \textbf{\bibinfo{volume}{80}},
  \bibinfo{pages}{490} (\bibinfo{year}{2020}).

\bibitem[{\citenamefont{Ovalle et~al.}(2021{\natexlab{a}})\citenamefont{Ovalle,
  Casadio, Contreras, and Sotomayor}}]{ovalle2021hairy}
\bibinfo{author}{\bibfnamefont{J.}~\bibnamefont{Ovalle}},
  \bibinfo{author}{\bibfnamefont{R.}~\bibnamefont{Casadio}},
  \bibinfo{author}{\bibfnamefont{E.}~\bibnamefont{Contreras}},
  \bibnamefont{and}
  \bibinfo{author}{\bibfnamefont{A.}~\bibnamefont{Sotomayor}},
  \bibinfo{journal}{Phys. Dark Universe} \textbf{\bibinfo{volume}{31}},
  \bibinfo{pages}{100744} (\bibinfo{year}{2021}{\natexlab{a}}).

\bibitem[{\citenamefont{Contreras et~al.}(2021)\citenamefont{Contreras, Ovalle,
  and Casadio}}]{contreras2021gravitational}
\bibinfo{author}{\bibfnamefont{E.}~\bibnamefont{Contreras}},
  \bibinfo{author}{\bibfnamefont{J.}~\bibnamefont{Ovalle}}, \bibnamefont{and}
  \bibinfo{author}{\bibfnamefont{R.}~\bibnamefont{Casadio}},
  \bibinfo{journal}{Phys. Rev. D} \textbf{\bibinfo{volume}{103}},
  \bibinfo{pages}{044020} (\bibinfo{year}{2021}).

\bibitem[{\citenamefont{Ovalle et~al.}(2021{\natexlab{b}})\citenamefont{Ovalle,
  Contreras, and Stuchlik}}]{ovalle2021kerr}
\bibinfo{author}{\bibfnamefont{J.}~\bibnamefont{Ovalle}},
  \bibinfo{author}{\bibfnamefont{E.}~\bibnamefont{Contreras}},
  \bibnamefont{and} \bibinfo{author}{\bibfnamefont{Z.}~\bibnamefont{Stuchlik}},
  \bibinfo{journal}{Phys. Rev. D} \textbf{\bibinfo{volume}{103}},
  \bibinfo{pages}{084016} (\bibinfo{year}{2021}{\natexlab{b}}).

\bibitem[{\citenamefont{Odintsov and Nojiri}(2006)}]{odintsov2006introduction}
\bibinfo{author}{\bibfnamefont{S.~D.} \bibnamefont{Odintsov}} \bibnamefont{and}
  \bibinfo{author}{\bibfnamefont{S.}~\bibnamefont{Nojiri}},
  \bibinfo{journal}{ECONF C} \textbf{\bibinfo{volume}{602061}},
  \bibinfo{pages}{06} (\bibinfo{year}{2006}).

\bibitem[{\citenamefont{Nojiri and Odintsov}(2007)}]{nojiri2007introduction}
\bibinfo{author}{\bibfnamefont{S.}~\bibnamefont{Nojiri}} \bibnamefont{and}
  \bibinfo{author}{\bibfnamefont{S.~D.} \bibnamefont{Odintsov}},
  \bibinfo{journal}{International Journal of Geometric Methods in Modern
  Physics} \textbf{\bibinfo{volume}{4}}, \bibinfo{pages}{115}
  (\bibinfo{year}{2007}).

\bibitem[{\citenamefont{Padmanabhan}(2008)}]{padmanabhan2008dark}
\bibinfo{author}{\bibfnamefont{T.}~\bibnamefont{Padmanabhan}},
  \bibinfo{journal}{Gen. Relativ. Gravit.} \textbf{\bibinfo{volume}{40}},
  \bibinfo{pages}{529} (\bibinfo{year}{2008}).

\bibitem[{\citenamefont{Clifton et~al.}(2012)\citenamefont{Clifton, Ferreira,
  Padilla, and Skordis}}]{clifton2012modified}
\bibinfo{author}{\bibfnamefont{T.}~\bibnamefont{Clifton}},
  \bibinfo{author}{\bibfnamefont{P.~G.} \bibnamefont{Ferreira}},
  \bibinfo{author}{\bibfnamefont{A.}~\bibnamefont{Padilla}}, \bibnamefont{and}
  \bibinfo{author}{\bibfnamefont{C.}~\bibnamefont{Skordis}},
  \bibinfo{journal}{Phys. Rep.} \textbf{\bibinfo{volume}{513}},
  \bibinfo{pages}{1} (\bibinfo{year}{2012}).

\bibitem[{\citenamefont{Capozziello and
  De~Laurentis}(2011)}]{capozziello2011extended}
\bibinfo{author}{\bibfnamefont{S.}~\bibnamefont{Capozziello}} \bibnamefont{and}
  \bibinfo{author}{\bibfnamefont{M.}~\bibnamefont{De~Laurentis}},
  \bibinfo{journal}{Phys. Rep.} \textbf{\bibinfo{volume}{509}},
  \bibinfo{pages}{167} (\bibinfo{year}{2011}).

\bibitem[{\citenamefont{Nojiri and Odintsov}(2011)}]{nojiri2011unified}
\bibinfo{author}{\bibfnamefont{S.}~\bibnamefont{Nojiri}} \bibnamefont{and}
  \bibinfo{author}{\bibfnamefont{S.~D.} \bibnamefont{Odintsov}},
  \bibinfo{journal}{Phys. Rep.} \textbf{\bibinfo{volume}{505}},
  \bibinfo{pages}{59} (\bibinfo{year}{2011}).

\bibitem[{\citenamefont{Sotiriou and Faraoni}(2010)}]{sotiriou2010f}
\bibinfo{author}{\bibfnamefont{T.~P.} \bibnamefont{Sotiriou}} \bibnamefont{and}
  \bibinfo{author}{\bibfnamefont{V.}~\bibnamefont{Faraoni}},
  \bibinfo{journal}{Rev. Mod. Phys.} \textbf{\bibinfo{volume}{82}},
  \bibinfo{pages}{451} (\bibinfo{year}{2010}).

\bibitem[{\citenamefont{De~Felice and Tsujikawa}(2010)}]{de2010f}
\bibinfo{author}{\bibfnamefont{A.}~\bibnamefont{De~Felice}} \bibnamefont{and}
  \bibinfo{author}{\bibfnamefont{S.}~\bibnamefont{Tsujikawa}},
  \bibinfo{journal}{Living Rev. Relativ.} \textbf{\bibinfo{volume}{13}},
  \bibinfo{pages}{1} (\bibinfo{year}{2010}).

\bibitem[{\citenamefont{Joyce et~al.}(2015)\citenamefont{Joyce, Jain, Khoury,
  and Trodden}}]{joyce2015beyond}
\bibinfo{author}{\bibfnamefont{A.}~\bibnamefont{Joyce}},
  \bibinfo{author}{\bibfnamefont{B.}~\bibnamefont{Jain}},
  \bibinfo{author}{\bibfnamefont{J.}~\bibnamefont{Khoury}}, \bibnamefont{and}
  \bibinfo{author}{\bibfnamefont{M.}~\bibnamefont{Trodden}},
  \bibinfo{journal}{Phys. Rep.} \textbf{\bibinfo{volume}{568}},
  \bibinfo{pages}{1} (\bibinfo{year}{2015}).

\bibitem[{\citenamefont{Nojiri et~al.}(2017)\citenamefont{Nojiri, Odintsov, and
  Oikonomou}}]{nojiri2017modified}
\bibinfo{author}{\bibfnamefont{S.}~\bibnamefont{Nojiri}},
  \bibinfo{author}{\bibfnamefont{S.~D.} \bibnamefont{Odintsov}},
  \bibnamefont{and} \bibinfo{author}{\bibfnamefont{V.~K.}
  \bibnamefont{Oikonomou}}, \bibinfo{journal}{Phys. Rep.}
  \textbf{\bibinfo{volume}{692}}, \bibinfo{pages}{1} (\bibinfo{year}{2017}).

\bibitem[{\citenamefont{Wagoner}(1970)}]{wagoner1970scalar}
\bibinfo{author}{\bibfnamefont{R.~V.} \bibnamefont{Wagoner}},
  \bibinfo{journal}{Phys. Rev. D} \textbf{\bibinfo{volume}{1}},
  \bibinfo{pages}{3209} (\bibinfo{year}{1970}).

\bibitem[{\citenamefont{Lovelock}(1972)}]{lovelock1972four}
\bibinfo{author}{\bibfnamefont{D.}~\bibnamefont{Lovelock}},
  \bibinfo{journal}{J. Math. Phys.} \textbf{\bibinfo{volume}{13}},
  \bibinfo{pages}{874} (\bibinfo{year}{1972}).

\bibitem[{\citenamefont{Ford}(1989)}]{ford1989inflation}
\bibinfo{author}{\bibfnamefont{L.}~\bibnamefont{Ford}}, \bibinfo{journal}{Phys.
  Rev. D} \textbf{\bibinfo{volume}{40}}, \bibinfo{pages}{967}
  (\bibinfo{year}{1989}).

\bibitem[{\citenamefont{Alcaraz et~al.}(2003)\citenamefont{Alcaraz, Cembranos,
  Dobado, and Maroto}}]{alcaraz2003limits}
\bibinfo{author}{\bibfnamefont{J.}~\bibnamefont{Alcaraz}},
  \bibinfo{author}{\bibfnamefont{J.~A.~R.} \bibnamefont{Cembranos}},
  \bibinfo{author}{\bibfnamefont{A.}~\bibnamefont{Dobado}}, \bibnamefont{and}
  \bibinfo{author}{\bibfnamefont{A.~L.} \bibnamefont{Maroto}},
  \bibinfo{journal}{Phys. Rev. D} \textbf{\bibinfo{volume}{67}},
  \bibinfo{pages}{075010} (\bibinfo{year}{2003}).

\bibitem[{\citenamefont{Nojiri and Odintsov}(2006)}]{nojiri2006modified}
\bibinfo{author}{\bibfnamefont{S.}~\bibnamefont{Nojiri}} \bibnamefont{and}
  \bibinfo{author}{\bibfnamefont{S.~D.} \bibnamefont{Odintsov}},
  \bibinfo{journal}{Phys. Rev. D} \textbf{\bibinfo{volume}{74}},
  \bibinfo{pages}{086005} (\bibinfo{year}{2006}).

\bibitem[{\citenamefont{Arai et~al.}(2023)\citenamefont{Arai, Aoki, Chinone,
  Kimura, Kobayashi, Miyatake, Yamauchi, Yokoyama, Akitsu, Hiramatsu
  et~al.}}]{arai2023cosmological}
\bibinfo{author}{\bibfnamefont{S.}~\bibnamefont{Arai}},
  \bibinfo{author}{\bibfnamefont{K.}~\bibnamefont{Aoki}},
  \bibinfo{author}{\bibfnamefont{Y.}~\bibnamefont{Chinone}},
  \bibinfo{author}{\bibfnamefont{R.}~\bibnamefont{Kimura}},
  \bibinfo{author}{\bibfnamefont{T.}~\bibnamefont{Kobayashi}},
  \bibinfo{author}{\bibfnamefont{H.}~\bibnamefont{Miyatake}},
  \bibinfo{author}{\bibfnamefont{D.}~\bibnamefont{Yamauchi}},
  \bibinfo{author}{\bibfnamefont{S.}~\bibnamefont{Yokoyama}},
  \bibinfo{author}{\bibfnamefont{K.}~\bibnamefont{Akitsu}},
  \bibinfo{author}{\bibfnamefont{T.}~\bibnamefont{Hiramatsu}},
  \bibnamefont{et~al.}, \bibinfo{journal}{Progress of Theoretical and
  Experimental Physics} \textbf{\bibinfo{volume}{2023}},
  \bibinfo{pages}{072E01} (\bibinfo{year}{2023}).

\bibitem[{\citenamefont{Bhatti et~al.}(2023)\citenamefont{Bhatti, Yousaf, and
  Yousaf}}]{bhatti2023novel}
\bibinfo{author}{\bibfnamefont{M.~Z.} \bibnamefont{Bhatti}},
  \bibinfo{author}{\bibfnamefont{M.}~\bibnamefont{Yousaf}}, \bibnamefont{and}
  \bibinfo{author}{\bibfnamefont{Z.}~\bibnamefont{Yousaf}},
  \bibinfo{journal}{Gen. Relativ. Gravit.} \textbf{\bibinfo{volume}{55}},
  \bibinfo{pages}{16} (\bibinfo{year}{2023}).

\bibitem[{\citenamefont{Kwon et~al.}(1986)\citenamefont{Kwon, Kim, Myung, Cho,
  and Park}}]{kwon1986stability}
\bibinfo{author}{\bibfnamefont{O.~J.} \bibnamefont{Kwon}},
  \bibinfo{author}{\bibfnamefont{Y.~D.} \bibnamefont{Kim}},
  \bibinfo{author}{\bibfnamefont{Y.~S.} \bibnamefont{Myung}},
  \bibinfo{author}{\bibfnamefont{B.~H.} \bibnamefont{Cho}}, \bibnamefont{and}
  \bibinfo{author}{\bibfnamefont{Y.~J.} \bibnamefont{Park}},
  \bibinfo{journal}{Phys. Rev. D} \textbf{\bibinfo{volume}{34}},
  \bibinfo{pages}{333} (\bibinfo{year}{1986}).

\bibitem[{\citenamefont{Shibata et~al.}(1994)\citenamefont{Shibata, Nakao, and
  Nakamura}}]{shibata1994scalar}
\bibinfo{author}{\bibfnamefont{M.}~\bibnamefont{Shibata}},
  \bibinfo{author}{\bibfnamefont{K.}~\bibnamefont{Nakao}}, \bibnamefont{and}
  \bibinfo{author}{\bibfnamefont{T.}~\bibnamefont{Nakamura}},
  \bibinfo{journal}{Phys. Rev. D} \textbf{\bibinfo{volume}{50}},
  \bibinfo{pages}{7304} (\bibinfo{year}{1994}).

\bibitem[{\citenamefont{Harada et~al.}(1997)\citenamefont{Harada, Chiba, Nakao,
  and Nakamura}}]{harada1997scalar}
\bibinfo{author}{\bibfnamefont{T.}~\bibnamefont{Harada}},
  \bibinfo{author}{\bibfnamefont{T.}~\bibnamefont{Chiba}},
  \bibinfo{author}{\bibfnamefont{K.-i.} \bibnamefont{Nakao}}, \bibnamefont{and}
  \bibinfo{author}{\bibfnamefont{T.}~\bibnamefont{Nakamura}},
  \bibinfo{journal}{Phys. Rev. D} \textbf{\bibinfo{volume}{55}},
  \bibinfo{pages}{2024} (\bibinfo{year}{1997}).

\bibitem[{\citenamefont{Jordan}(1938)}]{jordan1938empirischen}
\bibinfo{author}{\bibfnamefont{P.}~\bibnamefont{Jordan}},
  \bibinfo{journal}{Nature} \textbf{\bibinfo{volume}{26}}, \bibinfo{pages}{417}
  (\bibinfo{year}{1938}).

\bibitem[{\citenamefont{Brans and Dicke}(1961)}]{brans1961mach}
\bibinfo{author}{\bibfnamefont{C.}~\bibnamefont{Brans}} \bibnamefont{and}
  \bibinfo{author}{\bibfnamefont{R.~H.} \bibnamefont{Dicke}},
  \bibinfo{journal}{Phys. Rev.} \textbf{\bibinfo{volume}{124}},
  \bibinfo{pages}{925} (\bibinfo{year}{1961}).

\bibitem[{\citenamefont{Riess et~al.}(1998)}]{riess1998observational}
\bibinfo{author}{\bibfnamefont{A.~G.} \bibnamefont{Riess}}
  \bibnamefont{et~al.}, \bibinfo{journal}{Astron. J.}
  \textbf{\bibinfo{volume}{116}}, \bibinfo{pages}{1009} (\bibinfo{year}{1998}).

\bibitem[{\citenamefont{Banerjee and Pavon}(2001)}]{banerjee2001cosmic}
\bibinfo{author}{\bibfnamefont{N.}~\bibnamefont{Banerjee}} \bibnamefont{and}
  \bibinfo{author}{\bibfnamefont{D.}~\bibnamefont{Pavon}},
  \bibinfo{journal}{Phys. Rev. D} \textbf{\bibinfo{volume}{63}},
  \bibinfo{pages}{043504} (\bibinfo{year}{2001}).

\bibitem[{\citenamefont{Sen et~al.}(2001)\citenamefont{Sen, Sen, and
  Sethi}}]{sen2001dissipative}
\bibinfo{author}{\bibfnamefont{A.~A.} \bibnamefont{Sen}},
  \bibinfo{author}{\bibfnamefont{S.}~\bibnamefont{Sen}}, \bibnamefont{and}
  \bibinfo{author}{\bibfnamefont{S.}~\bibnamefont{Sethi}},
  \bibinfo{journal}{Phys. Rev. D} \textbf{\bibinfo{volume}{63}},
  \bibinfo{pages}{107501} (\bibinfo{year}{2001}).

\bibitem[{\citenamefont{Karimkhani and
  Khoadam-Mohammadi}(2019)}]{karimkhani2019hubble}
\bibinfo{author}{\bibfnamefont{E.}~\bibnamefont{Karimkhani}} \bibnamefont{and}
  \bibinfo{author}{\bibfnamefont{A.}~\bibnamefont{Khoadam-Mohammadi}},
  \bibinfo{journal}{Astrophys. Space Sci.} \textbf{\bibinfo{volume}{364}},
  \bibinfo{pages}{177} (\bibinfo{year}{2019}).

\bibitem[{\citenamefont{Singh and
  Sol{\`a}~Peracaula}(2021)}]{singh2021friedmann}
\bibinfo{author}{\bibfnamefont{C.}~\bibnamefont{Singh}} \bibnamefont{and}
  \bibinfo{author}{\bibfnamefont{J.}~\bibnamefont{Sol{\`a}~Peracaula}},
  \bibinfo{journal}{Eur. Phys. J. C} \textbf{\bibinfo{volume}{81}},
  \bibinfo{pages}{960} (\bibinfo{year}{2021}).

\bibitem[{\citenamefont{Ovalle et~al.}(2022)\citenamefont{Ovalle, Contreras,
  and Stuchlik}}]{ovalle2022energy}
\bibinfo{author}{\bibfnamefont{J.}~\bibnamefont{Ovalle}},
  \bibinfo{author}{\bibfnamefont{E.}~\bibnamefont{Contreras}},
  \bibnamefont{and} \bibinfo{author}{\bibfnamefont{Z.}~\bibnamefont{Stuchlik}},
  \bibinfo{journal}{Eur. Phys. J. C} \textbf{\bibinfo{volume}{82}},
  \bibinfo{pages}{211} (\bibinfo{year}{2022}).

\bibitem[{\citenamefont{Nordtvedt~Jr}(1970)}]{nordtvedt1970post}
\bibinfo{author}{\bibfnamefont{K.}~\bibnamefont{Nordtvedt~Jr}},
  \bibinfo{journal}{Astrophys. J.} \textbf{\bibinfo{volume}{161}},
  \bibinfo{pages}{1059} (\bibinfo{year}{1970}).

\bibitem[{\citenamefont{Ovalle}(2019)}]{ovalle2019decoupling}
\bibinfo{author}{\bibfnamefont{J.}~\bibnamefont{Ovalle}},
  \bibinfo{journal}{Phys. Lett. B} \textbf{\bibinfo{volume}{788}},
  \bibinfo{pages}{213} (\bibinfo{year}{2019}).

\bibitem[{\citenamefont{Ovalle and Casadio}(2020)}]{ovalle2020beyond}
\bibinfo{author}{\bibfnamefont{J.}~\bibnamefont{Ovalle}} \bibnamefont{and}
  \bibinfo{author}{\bibfnamefont{R.}~\bibnamefont{Casadio}},
  \emph{\bibinfo{title}{Beyond \textmd{E}instein Gravity: The Minimal Geometric
  Deformation Approach in the \textmd{B}rane-\textmd{W}orld}}
  (\bibinfo{publisher}{Springer Nature}, \bibinfo{year}{2020}).

\bibitem[{\citenamefont{Arias et~al.}(2020)\citenamefont{Arias, Tello-Ortiz,
  and Contreras}}]{arias2020extra}
\bibinfo{author}{\bibfnamefont{C.}~\bibnamefont{Arias}},
  \bibinfo{author}{\bibfnamefont{F.}~\bibnamefont{Tello-Ortiz}},
  \bibnamefont{and}
  \bibinfo{author}{\bibfnamefont{E.}~\bibnamefont{Contreras}},
  \bibinfo{journal}{Eur. Phys. J. C} \textbf{\bibinfo{volume}{80}},
  \bibinfo{pages}{463} (\bibinfo{year}{2020}).

\bibitem[{\citenamefont{Tello-Ortiz et~al.}(2020)\citenamefont{Tello-Ortiz,
  Maurya, and Gomez-Leyton}}]{tello2020class}
\bibinfo{author}{\bibfnamefont{F.}~\bibnamefont{Tello-Ortiz}},
  \bibinfo{author}{\bibfnamefont{S.~K.} \bibnamefont{Maurya}},
  \bibnamefont{and}
  \bibinfo{author}{\bibfnamefont{Y.}~\bibnamefont{Gomez-Leyton}},
  \bibinfo{journal}{Eur. Phys. J. C} \textbf{\bibinfo{volume}{80}},
  \bibinfo{pages}{324} (\bibinfo{year}{2020}).

\bibitem[{\citenamefont{Maurya and Nag}(2021)}]{maurya2021mgd}
\bibinfo{author}{\bibfnamefont{S.~K.} \bibnamefont{Maurya}} \bibnamefont{and}
  \bibinfo{author}{\bibfnamefont{R.}~\bibnamefont{Nag}}, \bibinfo{journal}{Eur.
  Phys. J. Plus} \textbf{\bibinfo{volume}{136}}, \bibinfo{pages}{679}
  (\bibinfo{year}{2021}).

\bibitem[{\citenamefont{Maurya et~al.}(2022)\citenamefont{Maurya, Singh,
  Govender, and Hansraj}}]{maurya2022gravitationally}
\bibinfo{author}{\bibfnamefont{S.~K.} \bibnamefont{Maurya}},
  \bibinfo{author}{\bibfnamefont{K.~N.} \bibnamefont{Singh}},
  \bibinfo{author}{\bibfnamefont{M.}~\bibnamefont{Govender}}, \bibnamefont{and}
  \bibinfo{author}{\bibfnamefont{S.}~\bibnamefont{Hansraj}},
  \bibinfo{journal}{Astrophys. J.} \textbf{\bibinfo{volume}{925}},
  \bibinfo{pages}{208} (\bibinfo{year}{2022}).

\bibitem[{\citenamefont{Israel}(1967)}]{israel1967nuovo}
\bibinfo{author}{\bibfnamefont{W.}~\bibnamefont{Israel}},
  \bibinfo{journal}{Erratum-ibid B} \textbf{\bibinfo{volume}{48}},
  \bibinfo{pages}{463} (\bibinfo{year}{1967}).

\bibitem[{\citenamefont{Darmois}(1927)}]{darmois1927memorial}
\bibinfo{author}{\bibfnamefont{G.}~\bibnamefont{Darmois}},
  \bibinfo{journal}{Fascicule XXV (Gauthier-Villars, Paris, 1927)}
  (\bibinfo{year}{1927}).

\bibitem[{\citenamefont{Israel}(1966)}]{israel1966singular}
\bibinfo{author}{\bibfnamefont{W.}~\bibnamefont{Israel}}, \bibinfo{journal}{Il
  Nuovo Cimento B (1965-1970)} \textbf{\bibinfo{volume}{44}},
  \bibinfo{pages}{1} (\bibinfo{year}{1966}).

\end{thebibliography}
\end{document}